\documentclass[singlecolumn]{revtex4}%
\usepackage{amsmath}
\usepackage{psfrag}
\usepackage{graphicx}

\newcommand{\vx}{\mathbf{x}}

\begin{document}

\title{ Spin-polarized Quantum Transport in Mesoscopic Conductors:\\
Computational Concepts and Physical Phenomena}

\author{Michael Wimmer, Matthias Scheid, and Klaus Richter}

\affiliation{Institut f{\"u}r Theoretische Physik, Universit{\"a}t
Regensburg, D-93040 Regensburg, Germany}

\maketitle

\tableofcontents

{
\vspace{1.2cm}
\em Glossary\\}

{\bf Aharonov Bohm effect}\\
The magnetic flux enclosed in between propagating quantum
mechanical waves shifts their relative phases as a result of the 
underlying electromagnetic vector potential. This gives rise 
to distinct oscillations in the magnetoconductance of a ring conductor.\\

{\bf Landauer-B\"uttiker formalism}\\
For phase-coherent quantum transport, the Landauer-B\"uttiker
formalism relates the conductance of a device to the transmission
probability of charge carriers.\\

{\bf Rashba- and Dresselhaus spin-orbit coupling} \\
Coupling of the spin degree of freedom to the orbital motion of charge carriers
due to structural or bulk inversion asymmetry in semiconductors.\\

{\bf Ratchets}\\
Devices that convert unbiased fluctuations or perturbations into 
directed motion.\\

{\bf Spintronics}\\
Extension of charge-based electronics in metals or
semiconductors by utilizing the spin degree of freedom of the 
charge carriers.\\

\section{Definition of the subject and its importance}

Mesoscopic conductors are electronic systems of sizes in
between nano- and micrometers, and often of reduced dimensionality.
In the phase-coherent regime at low temperatures, the conductance
of these devices is governed by quantum interference effects,
such as the Aharonov-Bohm effect and conductance fluctuations
as prominent examples. While first measurements of 
quantum charge transport date back to the 1980s, spin phenomena
in mesoscopic transport have moved only recently into the focus
of attention, as one branch of the field of spintronics.
The interplay between quantum coherence with confinement-, 
disorder- or interaction-effects gives rise to a variety of
unexpected spin phenomena in mesoscopic conductors and allows moreover 
to control and engineer the spin of the charge carriers: spin
interference is often the basis for spin-valves, -filters, -switches
or -pumps. Their underlying mechanisms may gain relevance on the 
way to possible future semiconductor-based spin devices.

A quantitative theoretical understanding of spin-dependent
mesoscopic transport calls for developing efficient and flexible 
numerical algorithms, including matrix-reordering techniques within
Green function approaches, which we will explain, review and employ.

\section{Introduction}
\label{sectionintroduction}

Charge and spin transport through phase-coherent conductors
of mesoscopic scales carry imprints of wave interference
as predominant and characteristic features: in the simplest 
case of a point contact, the conductance increases stepwise
with Fermi energy, reflecting the discrete number of quantized
open transverse channels contributing to transport; for more complex 
mesoscopic systems, such as ballistic quantum dots or diffusive 
conductors, the conductance typically wildly fluctuates, upon varying 
the Fermi energy or other parameters, around its classical mean value.
Among the different effects on charge transport, the Aharonov-Bohm (AB) effect 
represents possibly the most genuine interference phenomenon at the heart of mesoscopic 
physics: The magnetoconductance of a ring conductor coupled to two
leads exhibits distinct sinusoidal oscillations when monitored as a function 
of a perpendicular magnetic field threading the ring, with a period given 
by the magnetic flux quantum. As the AB signal stems from interfering waves 
travelling through the two different arms of the ring, it requires phase 
coherent wave functions extending over the ring typically on micron scales \cite{Datta2002}.
Hence the AB effect is frequently being used as a tool to 
investigate phase coherence and dephasing mechanisms of the orbital
part of the wave functions, while the spin degree of freedom 
was usually neglected. 

With rising interest in spin-dependent transport, 
the interplay of the electron spin and charge degree of freedom has 
been exploited in a variety of spin interference devices, to be discussed
below. Different types of couplings to the electron spin have been considered
for spin engineering in non-magnetic conductors. On the one hand this is 
possible through the Zeeman coupling to an externally applied magnetic field. 
Non-uniform $B$-fields with spatially varying direction are being employed 
to achieve a tailored spin dynamics, including the possibility for guided 
spin evolution or triggering spin flips. On the other hand, intrinsic spin-orbit 
(SO) interaction proves to be relevant in spin-dependent transport. It exists
in systems with bulk inversion asymmetry and/or structure
inversion asymmetry, e.g.\ due to the vertical confinement in semiconductor
heterostructures (Rashba SO coupling~\cite{Rashba1960,Bychkov1984}).

\begin{figure}[!tb]
\begin{center}
\includegraphics[width=0.45\linewidth]{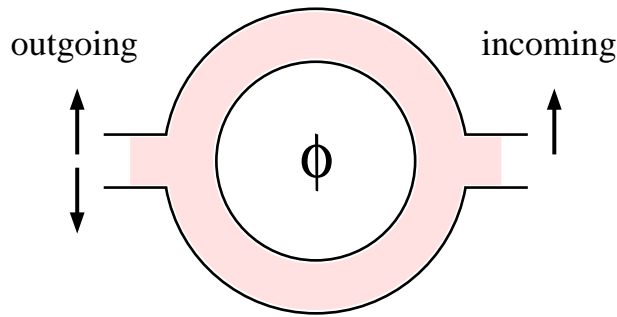}
\end{center}
\caption{Aharonov-Bohm physics with spin:
Two-dimensional Aharonov-Bohm ring of mean radius $r_0$ used for 
numerical calculations of the conductance presented in Fig.~\ref{spin-switch}.
An additional perpendicular magnetic field ${\bf B}$ generates a flux 
$\phi=\pi r_0^2 B$. 
The grey zone corresponds to the region subject to a finite Rashba coupling
switched adiabatically on and off in the leads (from Ref.~\cite{Frustaglia2004}).
}
\label{AB-ring}
\end{figure}

Among the mesoscopic spin interference systems considered in the literature,
ring geometries have again played an important role, opening up the field of 
spin-based AB physics (See Refs.~\cite{Hentschel2004,Cohen2007} for recent 
accounts including overviews over the literature).  This includes 
topics such as Berry phase-signatures in transport, both theoretically
\cite{LGB90,Stern92,AL-G93,QS94,FHR01,Frustaglia2004a,Nitta2007} and 
experimentally \cite{MHKWB98,YPS02,Koenig2006,Grbic2007}, spin-related
conductance modulation~\cite{Nitta1999,Malashukov1999,Frustaglia2004}, persistent 
currents \cite{LGB90,Splett2003}, spin Hall effect \cite{Souma2005}, 
spin filters~\cite{PFR03,Bellucci2007} and detectors~\cite{ID03}, 
and spin switching mechanisms \cite{FHR01,Hentschel2004}. 

To illustrate how the spin polarization can be tuned by exploiting (orbital and spin) 
interference in such an AB setup, we consider as an introductory example spin switching 
in a two-dimensional (2D) ballistic phase-coherent ring symmetrically coupled to two 
single-channel leads (Fig.~\ref{AB-ring}, Ref.~\cite{Frustaglia2004}). 
We assume Rashba SO coupling which is 
relevant in conductors laterally defined on GaAs- or InAs-based 
two-dimensional electron gases (2DEGs). Rashba SO coupling will be defined and
discussed in Sec.~\ref{spinfilter}. It can be viewed as the coupling
of the spin to a fictitious in-plane magnetic field directed perpendicular
to the electron momentum. Hence in a ring it points mainly in radial direction.
The strength of the SO field can be tuned by an external gate 
voltage~\cite{Nitta1997} allowing to control experimentally the spin evolution. 

\begin{figure}[!tb]
\begin{center}
\includegraphics[width=0.50\linewidth]{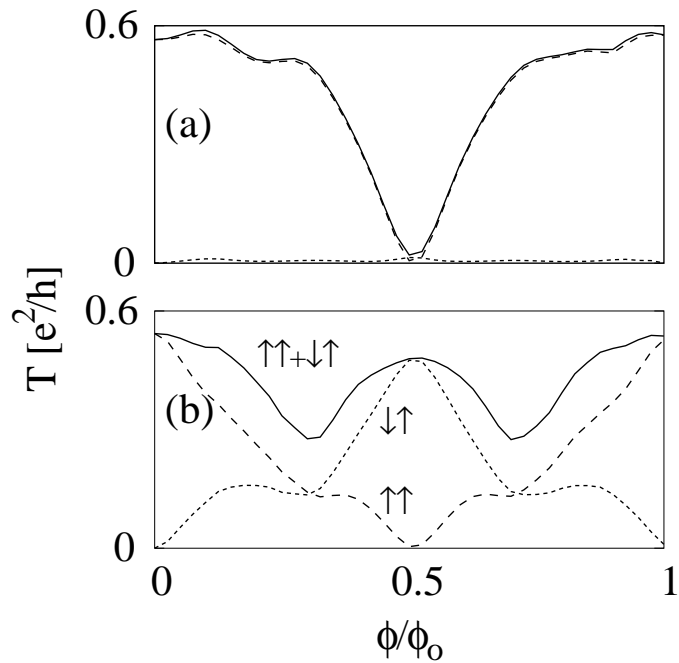}
\end{center}
\caption{Mesoscopic Aharonov-Bohm spin interference:
The conductance of spin-up polarized carriers entering a single-channel
two-dimensional Aharonov-Bohm ring (see Fig.~\ref{AB-ring}) is shown as a 
function of  flux $\phi=\pi r_0^2 B$ (in units of the flux quantum $\phi_0 = h c /e$)
through the ring in the presence of Rashba spin-orbit coupling. 
For panel (a) and (b) the scaled spin-orbit coupling strength takes the
values 0.2 and 1.0, expressed as the product $\omega_R T_0 / 2\pi$
of the precession frequency $\omega_R$ of the spin around
the effective spin-orbit magnetic field and the time $T_0$ for
 travelling of the electrons around the ring.
The overall conductance (solid line) is split into its
components $G^{\uparrow \uparrow}$ (dashed line) and $G^{\downarrow \uparrow}$ 
(dotted line).  Note in panel (b) the continuous change of the spin polarization,
related to $G^{\uparrow \uparrow} - G^{\downarrow \uparrow}$, with $\phi$ and the spin
switching at $\phi=\phi_0/2$
(adapted from Ref.~\cite{Frustaglia2004}).
}
\label{spin-switch}
\end{figure}

We assume {\it spin-polarized} spin-up electrons entering the ring from the 
right (see Fig.~\ref{AB-ring}). Fig.~\ref{spin-switch} displays
numerically computed (see Sec.~\ref{sectiontransport}) conductance traces 
as a function of the external magnetic flux $\phi$ for weak
and moderate Rashba strengths. The overall conductance is presented as a 
solid line, and its spin-resolved components, $G^{\uparrow \uparrow}$ and 
$G^{\downarrow \uparrow}$, corresponding to outgoing spin-up and -down 
channels, are shown as dashed and dotted lines, respectively.
In the weak SO coupling limit, Fig.~\ref{spin-switch}(a), the overall conductance 
(solid line) shows the usual AB oscillations of period $\phi_0$ and is dominated 
by $G^{\uparrow \uparrow}$ (dashed line). As expected for weak spin-coupling, 
the spin polarization is almost conserved during transport.
Interesting features appear for the case of moderate SO coupling
depicted in panel (b). There, both components,  $G^{\uparrow \uparrow}$
(dashed line) and $G^{\downarrow \uparrow}$ (dotted line), contribute 
similarly to the overall conductance (solid line). However, the spin polarization 
of the transmitted electrons varies as a function of the magnetic 
flux $\phi$:  $G^{\downarrow \uparrow}=0$ at $\phi=0$, 
while $G^{\uparrow \uparrow}=0$ at $\phi=\phi_0/2$.
Hence, for zero flux all transmitted carriers conserve their original
(incoming) spin-orientation, while for $\phi=\phi_0/2$ the transmitted
particles reverse their spin polarization. In other words: by tuning the
magnetic flux from $0$ to $\phi_0/2$ we can reverse the spin-polarization of
transmitted particles in a controlled way. Hence the AB ring with SO
coupling acts as a tunable spin-switch. This switching can be traced back to
constructive or destructive AB interference \cite{Hentschel2004}.

Switching a given spin polarization requires the generation
of spin-polarized particles in non-magnetic mesoscopic conductors
in the first place. Since spin injection from ferromagnets into a 
semiconductor remains problematic \cite{schmidt:2000}, alternative proposals 
have been made to achieve spin-polarized currents or spin accumulation
without magnets, which we will briefly review in Sec.~\ref{spinfilter}.
Among those are the spin Hall effect \cite{Sinova2004} and, in the context
of coherent mesoscopic transport, concepts for Zeeman- and SO-mediated 
adiabatic spin pumping and spin ratchets. 

For a recent account on spin phenomena in systems of reduced dimensions
see \cite{NJP-Focus}; for a review on the related field of 
magnetization dynamics and pumping 
in layered magnetic heterostructures see\cite{Tserkovnyak05}. 

A complete and quantitative understanding of spin phenomena in the mesoscopic 
realm requires computational approaches to quantum transport which 
also serve as reference calculations for analytical predictions
usually based on model assumptions. However, also numerical 
approaches cannot cope with the full many-body transport problem without
relying on approximations. Here we focus on mesoscopic conductors, i.e.\
systems with a considerable number of electrons, with strong coupling
to external leads. Then, a mean-field treatment is usually justified which
allows one to reduce the Hamiltonian to a single-particle problem with
an effective confinement potential resulting from a combination of external
and mean-field potentials. 

We further consider coherent transport close to
equilibrium at relatively low bias, excluding inelastic effects, such that 
the Landauer 
approach to transport is justified. However, even in this case brute-force
computational approaches quickly reach their limits:
Conductors at mesoscopic scales are typically characterized by extensions
which are (at least in one direction) much larger than the Fermi wavelength 
of the charge carriers, the shortest quantum scale involved.
This implies rapidly oscillating, complex and often irregular spinor wave functions 
extending throughout the systems which require for the quantum mechanical numerical 
solution either huge sets of basis functions or, in tight-binding 
approaches, the use of rather fine, preferentially adapted grids.
The strength of the wide-spread tight-binding approaches for transport
discussed and reviewed here lies in their flexibility and general applicability.
Moreover, tight-binding transport codes can be combined with density-functional 
(DFT) calculations for structure and electronic properties of the nano-conductors
by using the parameters computed within DFT as input. This approach
is frequently applied to transport in nano- or molecular electronics. 

The paper is organized as follows:
In the methodological Sec.~\ref{sectiontransport} we will first briefly summarize 
and provide the key relations for spin quantum transport within the Landauer framework. 
In Sec.~\ref{sectionreordering} we focus on and explain in some detail advanced computational 
concepts, making use of graph-theory, to implement powerful and flexible algorithms 
for tight-binding transport codes.
The numerical strength of the codes is demonstrated for ring-type geometries
showing that by efficient tight-binding implementation one can gain orders 
of magnitude in performance. In Secs.~\ref{spinfilter} and \ref{purespin} we employ 
these numerical schemes to address two important
aspects of spin-dependent transport, namely spin filtering and generating pure
spin currents. To this end we focus on laterally-confined 2D ballistic
nanostructures with Rashba SO interaction. We conclude with an outline of
future research directions in mesoscopic spin transport which we consider important.

\section{Numerical quantum transport}\label{sectiontransport}\label{algorithms}

\subsection{Landauer-B\"uttiker transport theory}

If the dimensions of a device get smaller than the phase coherence
length $l_\phi$ of charge carriers, classical transport theories 
are not valid any more. Instead, carrier dynamics is now governed by quantum
mechanics and the wave-like nature of particles becomes important. In general,
the conductance/resistance of such a device does not follow Ohm's law.

\begin{figure}[!tb]
\begin{center}
\includegraphics[width=0.65\linewidth]{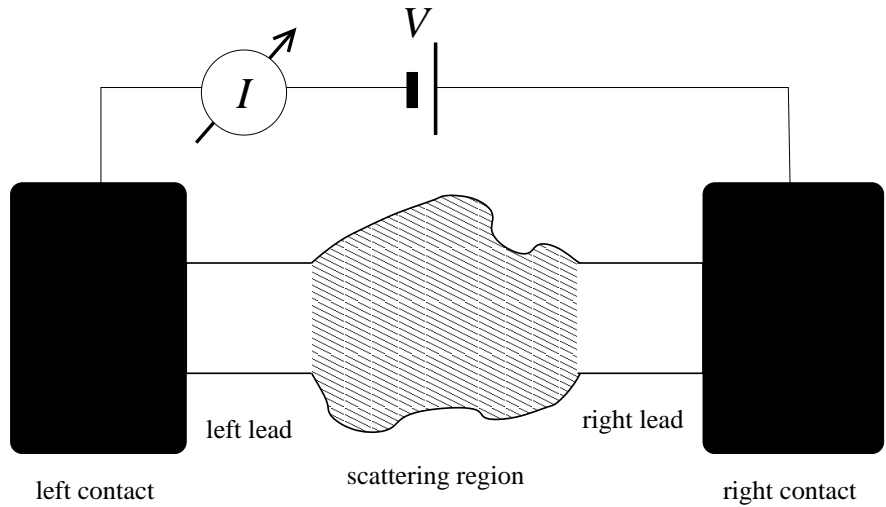}
\end{center}
\caption{Schematic view of a two-point measurement setup.}\label{landauerpic} 
\end{figure}

Consider a two-point measurement setup as shown in Fig.~\ref{landauerpic}: 
A scattering region is connected to large (phase-breaking) reservoirs by 
leads. The leads are assumed to be perfect and infinitely long to 
define asymptotic eigenstates $\phi_{n,\sigma}(y) e^{\pm ikx}$ at energy $E$, 
where $n$ is the quantum number of transverse confinement --- also called the
\emph{channel} number --- and $\sigma$ is the spin index. The total
scattering eigenstate $\psi_{n,\sigma}$ originating from channel $n$
with spin $\sigma$ in the left lead is, within the lead region, given by 
\begin{equation}
\psi_{n,\sigma}(\vx)=
\begin{cases}
\phi_{n,\sigma}(y) e^{i k x} + \sum_{m} r_{m,\sigma';n,\sigma}\;\, \phi_{m,\sigma'}(y) e^{-ikx}& \text{for $\vx$ in left lead}\\
\sum_{m} t_{m,\sigma';n,\sigma}\;\, \phi_{m,\sigma'}(y) e^{ikx}& \text{for $\vx$ in right lead}
\end{cases}\;,
\end{equation}
and obeys the stationary Schr\"odinger equation 
$H\psi_{n,\sigma}=E\psi_{n,\sigma}$. The conductance $G$ in linear response
can then be calculated within 
the \emph{Landauer-B\"uttiker} 
formalism~\cite{Landauer1957,Buttiker1985,Stone1988} 
(for tutorials see~\cite{Datta2002, Ferry2001}):
\begin{equation}\label{landauerformula}
G=\frac{e^2}{h} \sum_{n,m} \sum_{\sigma,\sigma'}\; T_{m,\sigma';n,\sigma} =
\frac{e^2}{h}\; T_\mathrm{C}\;,
\end{equation}
where $T_{m,\sigma';n,\sigma}
=\left|t_{m,\sigma';n,\sigma}\right|^2$ is given
by the squared transmission amplitudes of the 
scattering states $\psi_{n,\sigma}$.
The fraction $\frac{e^2}{h}$ is called the conductance quantum.
The \emph{scattering matrix} $S_{m,\sigma';n,\sigma}$ is a useful
definition that combines reflection
and transmission amplitudes for both leads into a single matrix. In this
notation the index
$n,\sigma$ then also contains information about the respective lead. 

The problem of calculating the conductance $G$ is thus reduced to calculating
the scattering eigenstates $\psi_{n,\sigma}$. Alternatively, the scattering 
amplitudes $ r_{m,\sigma';n,\sigma}$ and $t_{m,\sigma';n,\sigma}$ can also be 
derived from the retarded Greens function $G^\mathrm{R}(\vx,\vx')$ 
of the system. The retarded Greens function for a given energy $E$ 
obeys the equation 
\begin{equation}
(E-H+i\eta)\;G^\mathrm{R}(\vx,\vx') = \delta(\vx-\vx')\;,
\end{equation}
where $H$ is the Hamiltonian of the system and $\eta$ an infinitesimally small
number. Formally this equation can be solved as 
\begin{equation}\label{formalGR}
G^\mathrm{R}=(E-H +i\eta)^{-1}\;.
\end{equation}
The transmission and reflection amplitudes are then given by the 
\emph{Fisher-Lee} relation~\cite{Fisher1981}: 
\begin{equation}\label{fisherleetransmission}
t_{m,\sigma';n,\sigma}=-i \hbar \sqrt{v_{m} v_n}\, \int_{C_\mathrm{R}} dy 
\int_{C_\mathrm{L}} dy'
\phi_{m,\sigma'}(y)\,G^\mathrm{R}(\vx, \vx')\,\phi_{n,\sigma}(y')\;,
\end{equation}
\begin{equation}\label{fisherleereflection}
r_{m,\sigma';n,\sigma}=\delta_{mn}\delta_{\sigma'\sigma}-i \hbar \sqrt{v_{m} v_n}\, \int_{C_\mathrm{L}} dy \int_{C_\mathrm{L}} dy'
\phi_{m,\sigma'}(y)\,G^\mathrm{R}(\vx, \vx')\,\phi_{n,\sigma}(y')\;,
\end{equation}
where $v_n$ denotes the velocity of channel $n$ and the integration runs
over the cross-section $C_\mathrm{L}$ ($C_\mathrm{R}$) of the 
left (right) lead. Equations (\ref{fisherleetransmission}) and 
(\ref{fisherleereflection}) are valid only for leads without magnetic fields 
and no spin-orbit interaction. Baranger and Stone~\cite{Baranger1989} have
extended the formalism to also account for arbitrary 
magnetic fields in the leads, and their description can also be applied to
finite spin-orbit interaction.  

\subsection{Tight-binding representation of the Hamiltonian}

Except for very simple geometries, the scattering problem cannot be solved
analytically. Therefore, the use of computers for a numerical
solution of the scattering problem is very often the method of choice. 
However, the related stationary Schr\"odinger equation $H\psi=E\psi$ is a differential 
equation with continuous degrees of freedom 
that are difficult to treat on a computer. 
In general, a numerical solution is thus only attempted within a 
\emph{discrete} basis set which converts the differential equation
into a matrix equation. 

The method of finite differences is a very simple and yet powerful way 
to introduce such a discrete basis set and has been applied to the 
Schr\"odinger equation already as early as 
1934~\cite{Kimball1934, Pauling1935} (for an introduction 
see~\cite{Datta2002, Ferry2001}). Here we illustrate, as an example, the 
application of the method for a simple one-dimensional 
effective mass Hamiltonian including a potential
\begin{equation}\label{simplehamiltonian}
H=-\frac{\hbar^2}{2 m} \frac{d^2}{dx^2} + V(x)\;.
\end{equation}

\begin{figure}
\includegraphics[width=0.75\linewidth]{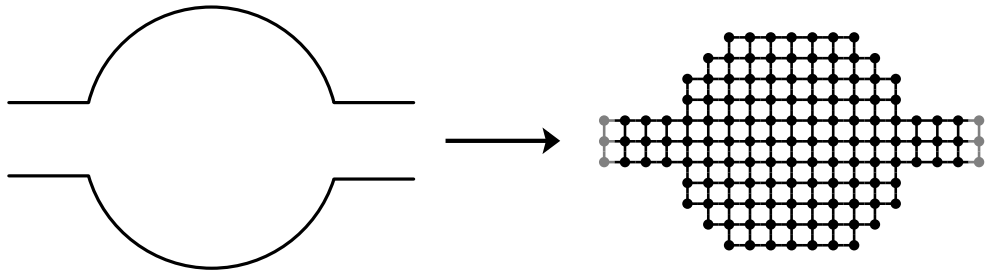}
\caption{Discretizing a continuous region on a square grid.}
\label{discretization}
\end{figure}

In the method of finite differences, space is approximated by a grid of
discrete lattice points spaced equidistantly with lattice constant $a$. 
For the 2D case, usually a square grid 
is used as depicted in Fig.~\ref{discretization}. Using the Taylor
expansion of the wave function $\psi$ we can write
\begin{equation}\label{taylor1}
\psi(x+a)=\psi(x)+\psi'(x) a +\frac{1}{2}\psi''(x) a^2 + 
\frac{1}{6} \psi^{(3)}(x) a^3 + \frac{1}{24} \psi^{(4)}(x) a^4 + \dots\;,
\end{equation}
\begin{equation}\label{taylor2}
\psi(x-a)=\psi(x)-\psi'(x) a +\frac{1}{2}\psi''(x) a^2 -
\frac{1}{6} \psi^{(3)}(x) a^3 + \frac{1}{24} \psi^{(4)}(x) a^4 + \dots\;.
\end{equation}
Adding Eqs.~(\ref{taylor1}) and~(\ref{taylor2}), we arrive at an expression for
the second derivative of the wave function in terms of values
of the wave function on the grid,
\begin{equation}
\frac{d^2}{dx^2}\psi(x)=\frac{1}{a^2} \left(\psi(x+a)+\psi(x-a)-2\psi(x)\right)
+\mathcal{O}(a^2)\;,
\end{equation}
valid up to second order in the lattice spacing $a$. The differential
equation 
\begin{equation}
-\frac{\hbar^2}{2 m} \frac{d^2}{dx^2}\psi(x) + V(x)\psi(x) = E\psi(x)
\end{equation}
is thus replaced by a set of difference equations 
\begin{equation} 
-\frac{\hbar^2}{2 m a^2} \left(\psi(x_{i+1})+\psi(x_{i-1})-2\psi(x_i)\right)
+V(x) \psi(x_i) = E \psi(x_i)\;,
\end{equation} 
where $x_{i\pm 1}=x_i\pm a$, yielding the 
tight-binding representation of the Hamiltonian:
\begin{equation}
H=\sum_{x_i} -\frac{\hbar^2}{2 m a^2} \left(\left|x_i\right>\left<x_{i+1}\right| + \text{h.c.} \right)
+\left(2\frac{\hbar^2}{2 m a^2}+V(x_i)\right) 
\left|x_i\right>\left<x_i\right|\,.
\end{equation}
Here, $\left|x_i\right>$ denotes a state localized at grid point $x_i$. 

\begin{figure}
\begin{center}
\includegraphics[width=0.35\linewidth]{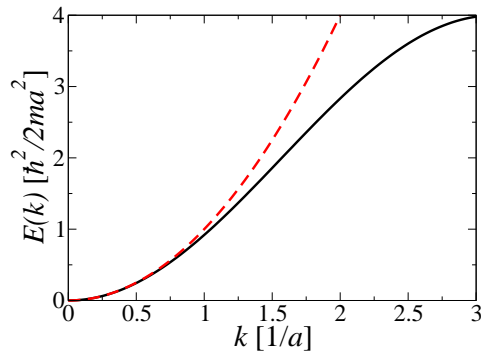}
\end{center}
\caption{Energy spectrum $E(k)$ for the continuous Schr\"odinger equation
(red dashed line) and the tight-binding approximation (black solid line)}
\label{FDvalidity}
\end{figure}

In principle, the quality of the finite differences approximation can be 
improved up to a desired precision by reducing the lattice spacing $a$.
However, since this leads to a larger problem size, the minimum lattice
spacing achievable is set by the available computing time and memory. Thus,
one must keep an eye on the validity of the finite differences approximation.
In Fig.~\ref{FDvalidity} we show the energy spectrum $E(k)$ for the
continuous one-dimensional Schr\"odinger equation and the 
tight-binding (finite differences)
approximation. The tight-binding approximation only holds for $k a\ll 1$
and $E\ll\frac{\hbar^2}{2ma^2}$, and it does not make sense to consider the
whole energy spectrum given by the tight-binding band width. The 
method of finite differences presented here can straight-forwardly be
applied to more complex Hamiltonians, including for example
spin-orbit interactions~\cite{Frustaglia2004}, and will be 
later used to calculate transport phenomena including spin in 
sections \ref{spinfilter} and \ref{purespin}.

A tight-binding representation of the Hamiltonian can also be 
obtained by applying the finite element method~\cite{Havu2004}. Furthermore
tight-binding Hamiltonians are also used in treatments beyond
the effective mass approximation, such as from atomic orbitals 
in empirical tight-binding models~\cite{Bowen1997,Sanvito1999,Luisier2006}, 
or from orbitals of the Kohn-Sham equations within 
DFT~\cite{Brandbyge2002,DiCarlo2006,Rocha2006}.

\subsection{Numerical algorithms}

Within the tight-binding approximation the Hamiltonian $H$ can be
represented by a matrix, even though this matrix is still infinite
as the leads are infinitely long. 
However, the infinite matrix problem can be reduced to a finite problem 
by partitioning the system into three isolated parts: left lead, scattering
region, and right lead. The Hamiltonian
then reads
\begin{equation}
H=\left(\begin{array}{ccc}
H_\mathrm{L}&V_\mathrm{LS}&0\\
V_\mathrm{SL}&H_\mathrm{S}&V_\mathrm{SR}\\
0&V_\mathrm{RS}&H_\mathrm{R}
\end{array}
\right)\;,
\end{equation} 
where $H_\mathrm{L(R)}$ is the (infinite) Hamiltonian of the left (right) lead, 
$H_S$ is the Hamiltonian of the scattering region and of finite size. Since 
the leads are always chosen such that asymptotic eigenstates can be defined,
the Hamiltonian of the isolated leads must contain some periodicity that 
facilitates calculating their Greens functions $g^\mathrm{R}_\mathrm{L,R}$. 
This can be done analytically for simple systems~\cite{Datta2002,Ferry2001}, 
for more complex situations the Greens function can be calculated numerically
either by iteration~\cite{LopezSancho1984,LopezSancho1985} or 
semianalytical formulas~\cite{Sanvito1999,Krstic2002,Rocha2006}. 
Introducing the retarded self-energy $\Sigma^\mathrm{R}=\sum_{i=L,R}\; 
V_{\mathrm{S}i}\,g^\mathrm{R}_i\,V_{i\mathrm{S}}$,~\cite{Datta2002,Ferry2001} the
Greens function $G_\mathrm{S}$ of the scattering region is given by
\begin{equation}\label{directinversion}
G^\mathrm{R}_\mathrm{S}=\left(E-H-\Sigma^\mathrm{R}\right)^{-1}\;,
\end{equation}
reminiscent of Eq.~(\ref{formalGR}) but with an effective Hamiltonian
$H+\Sigma^\mathrm{R}$.

\begin{figure}
\begin{center}
\includegraphics[width=0.4\linewidth]{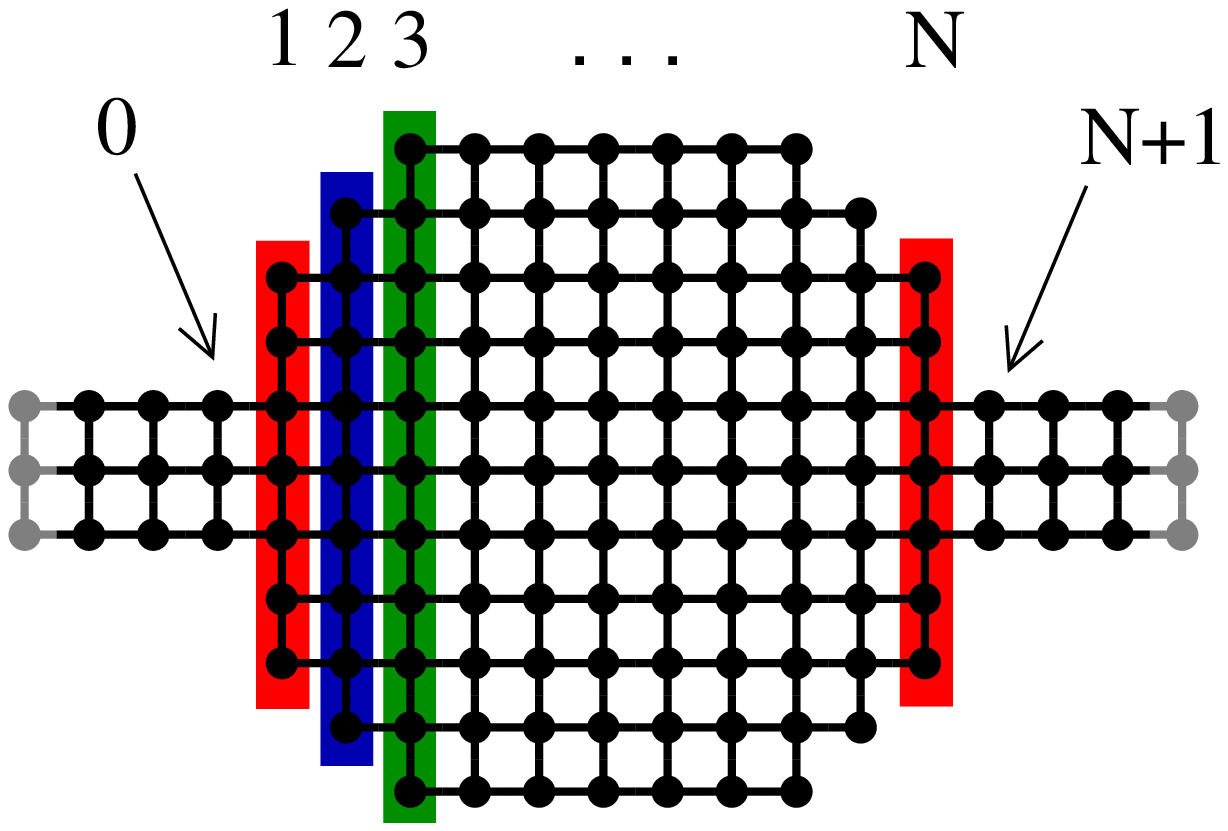}
\end{center}
\caption{Block-tridiagonal matrix form arising in the method of finite differences.
Grid points with the same $x$-coordinate are placed into the same block.}\label{BTDform}
\end{figure}

The original infinite-dimensional problem has thus been reduced to a finite size matrix 
problem that can, in principle, be solved straight-forwardly on a computer. However, 
for any but rather small problems, the computational task of the direct inversion 
in Eq.~(\ref{directinversion}) is prohibitive. Therefore, many algorithms make 
use of the \emph{sparsity} of the Hamiltonian matrix in tight-binding 
representation  --- in particular they employ the property that this matrix can be 
written in block-tridiagonal form
\begin{equation}
H=\left(
\begin{array}{@{\extracolsep{4mm}}cccccc@{\extracolsep{0mm}}ccccc}
\ddots&&&\\
&H_\mathrm{L}&V_\mathrm{L}&\\
&V^\dagger_\mathrm{L}&H_\mathrm{L}&H_{01}&&&\ddots\\
&&H_{10}&H_{11}&H_{12}&&&0\\
&&&H_{21}&H_{22}&H_{23}&&&\ddots\\
&&&&H_{32}&\ddots\\
&&\ddots&&&&\ddots&H_{N-1 N}\\
&&&0&&&H_{N N-1}&H_{NN}&H_{N N+1}\\
&&&&\ddots&&&H_{N+1 N}&H_\mathrm{R}&V_\mathrm{R}\\
&&&&&&&&V^\dagger_\mathrm{R}&H_\mathrm{R}\\
&&&&&&&&&&\ddots
\end{array}\right)\; .
\end{equation}
Here the index $\mathrm{L}$ ($\mathrm{R}$) denotes the blocks in the left (right) lead,
$1 \dots N$ the blocks within the scattering region, and $0$, ($N+1$) the first block in
the left (right) lead. Such a form arises naturally in the method of finite differences, 
when grid points are grouped into vertical slices according to their $x$-coordinates, 
as shown in Fig.~\ref{BTDform}, but also applies to any other sparse tight-binding 
Hamiltonians. 

The block-tridiagonal form of the Hamiltonian is the foundation
of several quantum transport algorithms, together with the fact that, according
to Eqs.~(\ref{fisherleetransmission}) and (\ref{fisherleereflection}), only the blocks
$G^\mathrm{R}_{N+1\,0}$ and $G^\mathrm{R}_{00}$ of the Greens function $G^\mathrm{R}$ are
needed for the calculation of transmission and reflection probabilities.
The transfer matrix approach applies naturally to block-tridiagonal
Hamiltonians, but becomes unstable for larger systems. However, a stabilized version
has been developed by Usuki \emph{et al.}~\cite{Usuki1994,Usuki1995}. 
In the decimation technique~\cite{Lambert1980,Leadbeater1998},
the Hamiltonian of the scattering region is replaced by an effective Hamiltonian
between the two leads by eliminating internal degrees of freedom. The contact block reduction 
method~\cite{Mamaluy2005} calculates the full Greens function of the system using a limited set of
eigenstates. The recursive Greens function (RGF) technique~\cite{Thouless1981,Lee1981,MacKinnon1985}
uses Dyson's equation to build up the system's Greens function block by block. It has also been adapted
to Hall geometries with four terminals~\cite{Baranger1991} and
to calculate non-equilibrium densities~\cite{Lake1997,Lassl2007}. Furthermore, the RGF algorithm
has been formulated to be suitable for parallel computing~\cite{Drouvelis2006},
and the modular recursive Greens function (MRGF) method is an extension to take advantage
of special geometries, such as circles or rectangles~\cite{Rotter2000,Rotter2003}.

\begin{figure}
\begin{center}
\includegraphics[width=0.6\linewidth]{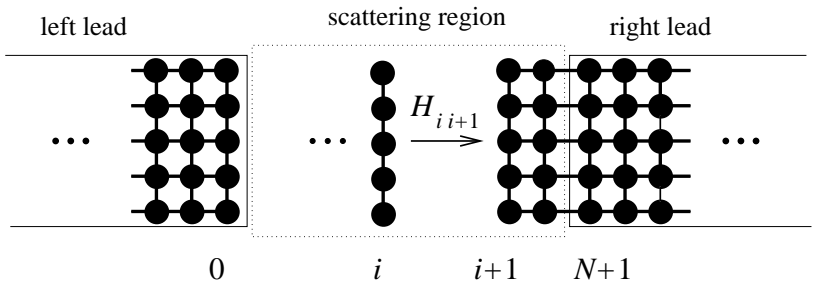}
\end{center}
\caption{Schematic view of the recursive Greens function algorithm: the
system is built up adding block by block.}\label{RGFalgo}
\end{figure}

Here we give a simple form of the RGF algorithm as described in~\cite{MacKinnon1985, Ferry2001}. 
The RGF method makes use of Dyson's equation
\begin{equation}\label{dysonequation}
G=G_0+G_0 V G
\end{equation}
to add successively blocks to the system, as depicted in Fig.~\ref{RGFalgo}. Let
$G^{\mathrm{R},(i)}$ denote the Greens function for the system containing all blocks
$\geq i$. Then, at energy $E$, the Greens function $G^{\mathrm{R},(i)}$ is related to $G^{\mathrm{R},(i+1)}$ by
\begin{equation} 
G^{\mathrm{R},(i)}_{i\,i}=\left(E-H_{i\,i}-
H_{i\,i+1}\;G^{\mathrm{R},(i+1)}_{i+1\,i+1}\;H_{i+1\,i}\right)^{-1}\;\;\text{and}
\end{equation}
\begin{equation} 
G^{\mathrm{R},(i)}_{N+1\,i}=G^{\mathrm{R},(i+1)}_{N+1\,i+1}\;H_{i+1\,i}\;G^{\mathrm{R},(i)}_{i\,i}\;.
\end{equation}
Starting from $G^{\mathrm{R},(N+1)}_{N+1\,N+1}=g^{\mathrm{R}}_\mathrm{R}$, the surface Greens function of the right
lead, $N$ slices are added recursively, until $G^{\mathrm{R},(1)}$ has been calculated. The 
blocks of the Greens function of the full system necessary for transport are then given by
\begin{equation}  
G^{\mathrm{R}}_{00}=\left(\left(g^\mathrm{R}_\mathrm{L}\right)^{-1}-\;H_{01}\;G^{\mathrm{R},(1)}_{11}\;H_{10}\right)^{-1}\;\;
\text{and}
\end{equation}
\begin{equation}
G^{\mathrm{R}}_{N+1\,0}=G^{\mathrm{R},(1)}_{N+1\,1}\;H_{10}\;G^{\mathrm{R}}_{00}\;,
\end{equation}
where $g_\mathrm{L}^\mathrm{R}$ is the surface Greens function of the left lead.

Each step of the algorithm performs inversions and matrix multiplications with matrices
of size $M_i$. Since the computational complexity of matrix inversion and multiplications
scales as $M_i^3$, the complexity of the RGF algorithm is $\propto \sum_{i=0}^{N+1} M_i^3$. Thus, it
scales linearly with the ``length'' N, and cubically with the ``width'' $M_i$ of the system.

While for certain geometries the RGF algorithm cannot compete with more specialized algorithms
such as MRGF, it is very versatile and easily adapted to many situations, and is thus our
method of choice. In the next section we will discuss matrix reordering techniques
that improve the runtime of the RGF algorithm considerably and allow the treatment 
of arbitrary systems.

\section{Matrix reordering strategies for quantum transport}\label{sectionreordering}

\subsection{Graph-theoretical approaches to matrix reordering}

\begin{figure}
\begin{center}
\includegraphics[width=0.6\linewidth]{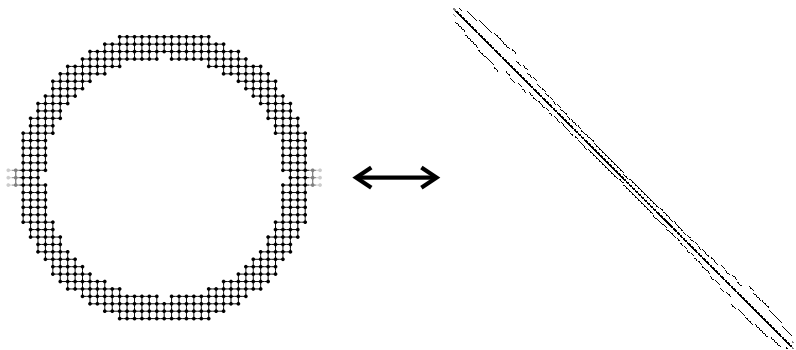}
\end{center}
\caption{Example of a sparsity structure of a matrix associated with a finite difference grid. Black
dots mark non-zero entries. The picture of the finite difference grid
can also be interpreted as a graphical representation of a graph.}\label{naturalsparsematrix}
\end{figure}

As shown in the previous section, the structure of a Hamiltonian matrix $H$ does
influence the runtime of the RGF algorithm. Thus, the runtime of
the algorithm can potentially be improved by reordering the matrix with a permutation $P$,
\begin{equation}
H'=P\;H\;P^{-1}\;.
\end{equation}
At first glance, such an effort may seem pointless: For example, the block-tridiagonal
structure naturally associated with a finite difference grid (as discussed in
section~\ref{sectiontransport}) leads to a sparse matrix with a small bandwidth, as shown
in Fig.~\ref{naturalsparsematrix}. However, as we show below, the choice of a suitable
permutation $P$ can still lead to a significant speed-up of the RGF algorithm.

For this we define a weight $w(H)$ associated with a given matrix $H$ as 
\begin{equation}
w(H)=\sum_{i=0}^{N+1} M_i^3\qquad\text{where $M_i$ is the size of block $H_{ii}$}\;.
\end{equation}
Optimizing the matrix for the RGF algorithm is then equivalent to minimizing the 
weight $w(H)$. Since $\sum_{i=0}^{N+1} M_i = N_\mathrm{grid}$, where $N_\mathrm{grid}$ 
is the total number of grid points, $w(H)$ is minimal, if all $M_i$ are equal, and 
$M_i=N_\mathrm{grid}/(N+2)$. Therefore, a matrix tends to have small weight, if the number $N$ of
blocks is large, and all blocks are equally sized. The reordering problem
of the matrix $H$ is thus summarized as follows:

{\bf Matrix reordering problem:}
Find a reordered matrix $H'$ such, that
\begin{enumerate}
\item $H'_{00}$ and $H'_{N+1N+1}$ are blocks given by the left and
right leads (as required by the RGF algorithm),
\item $H'$ is block-tridiagonal ($H'_{ij}\neq 0$, iff $j=i+1,i,i-1$),
\item the number $N$ of blocks is as large as possible, and
all blocks are equally sized.
\end{enumerate}
These requirements define the optimization problem of reordering
the matrix $H$. Usually, in such optimization problems finding
the best solution deterministically is prohibitively expensive, and one 
has to resort to heuristic strategies. 

In order to do that, we reformulate our matrix problem using 
concepts from graph theory. A \emph{graph} $G$ is a
pair $G=(V, E)$, where $V$ is a set of \emph{vertices} $i$, and 
$E$ a set of pairs of vertices $(i,j) \in V\times V $. Such a pair is called
an \emph{edge}. A graph is called \emph{undirected}, if for every edge
$(i,j)\in E$ also $(j,i)\in E$. Two vertices $i$ and $j$  are called \emph{adjacent}, if
$(i,j)\in E$. A graph can be visualized by drawing
dots for each vertex $i$ and lines connecting these dots for every 
edge $(i,j)$. It should be noted, that all the
pictures of finite difference grids shown so far can directly be interpreted 
as graphs .
There is a natural one-to-one correspondence between graphs and the structure
of sparse matrices. For a given $n\times n$ matrix $H$, we define a graph
$G=(V,E)$ with $V=\{1,\dots, n\}$ and $(i,j)\in E$ iff $H_{ij} \neq 0$. Thus, 
the symmetric zero--nonzero structure of Hermitian matrices, 
as considered in quantum transport, leads to associated undirected graphs.
An example of such a correspondence between a graph and a matrix
has already been shown in Fig.~\ref{naturalsparsematrix}. With respect
to matrices, graphs are also very convenient for 
storing and handling sparse matrix data structures on a computer. 

In terms of graph theory, 
matrix reordering corresponds to renumbering the vertices of a graph.
Since we are only interested in reordering the matrix
in terms of matrix blocks (the order within a block should not
matter too much), we define a \emph{partitioning} of $G$
as a set $\{V_i\}$ of disjoint subsets $V_i \subset V$ such that
$\bigcup_{i} V_i = V$ and $V_i \cap V_j =\emptyset$ for $i\neq j$. 
Using these concepts, we can now reformulate the original matrix reordering
problem into a graph partitioning problem:

{\bf Graph partitioning problem:} Find a partitioning $\{V_0, \dots, V_{N+1}\}$
of $G$ such that:
\begin{enumerate}
\item $V_0$ and $V_{N+1}$ contain the vertices belonging to left and right leads,
\item there are edges between $V_i$ and $V_j$ iff $j=i+1,i,i-1$ (block-tridiagonality),
\item the number of sets $V_i$ is as large as possible and all sets $V_i$ have the
same cardinality. Such a partitioning with all $V_i$ equally sized
is called \emph{balanced}. 
\end{enumerate} 

A partitioning that meets requirement 2 is called a \emph{level set} with
levels $V_i$ and appears
commonly as an intermediate step in algorithms for bandwidth reduction of
a matrix~\cite{Cuthill1969,George1971,Liu1976,Gibbs1976}. These algorithms seek
to find a level set of minimal width, i.~e. $\max_{i=0\dots N+1} |V_i|$ as small as possible
which is equivalent to requirement 3. The major difference between our graph partitioning
problem and the bandwidth reduction algorithms is requirement 1: In our case 
the start and end blocks are given by the geometry of the system, whereas in the 
bandwidth reduction methods these can be chosen freely. The implications of this
difference will be discussed below. 

The bandwidth reduction algorithms start with the observation that
a \emph{breadth first search} (BFS) starting from any vertex in the graph creates
a level set: In our situation, the BFS starts from level $V_0$ of the left lead. 
Then, successively for $i=0,1,2,\dots$, all vertices adjacent to $V_i$ that have 
not been assigned to a level yet are placed in $V_{i+1}$.
This construction ensures that each level $V_i$ only has
edges connecting to vertices in $V_{i+1,i,i-1}$ and thus ensures block-tridiagonality.
The search stops, once a vertex adjacent to the right lead is encountered, and 
all unassigned vertices are placed into the last level $V_N$. The number of
BFS steps determines therefore the maximum number $N$ of blocks and is
related to the minimum distance between left and right lead in the graph. However, 
this construction of the level set also suffers from a serious problem: Depending on the
distance between the leads, the last level $V_N$ can potentially contain a large 
number of vertices leading to a very unbalanced partitioning. In the bandwidth reduction
methods, the first and the last vertex are chosen to have (to a good approximation)
maximum distance, and thus this problem does not occur there. Hence, conventional 
bandwidth reduction algorithms can only be applied to quantum transport
problems, if the leads are --- in terms of the underlying graph --- furthest apart.

In this study, we will consider two reordering algorithms: First, the Gibbs-Poole-Stockmeyer
(GPS) algorithm~\cite{Gibbs1976}, a widely used bandwidth reduction algorithm. The GPS algorithm
combines the level sets originating from a BFS from both left and right lead to give
an optimized level set. Due to the limitations discussed above, it can only be used
efficiently for systems with leads far apart. To overcome this difficulty partly, 
we also propose a second algorithm, later referred to as
\emph{BFS partitioning}: The system is bisected recursively by means
of a simultaneous BFS from left and right leads. In a bisection process, vertices
that are closer (as given by the number of BFS steps) to the left (right) lead
are placed into the left (right) level. The resulting two levels are then further
bisected recursively until the final level set has been constructed. This algorithm
tries to avoid an unbalanced partitioning, as every step
tries to create a balanced bisection. We have found this global approach --- as opposed to the 
local approach in the BFS --- to yield balanced partitionings for systems where 
there are only few local minima in the weight $w(H)$. For general systems, a more
sophisticated method should be used~\cite{WimmerRichterunpublished}.     

Obviously, reordering the matrix will only improve the runtime of the
RGF algorithm, if the overhead of the reordering process is small 
compared to the actual transport calculation. Because of this reason,
applying general optimization algorithms to the matrix reordering problem 
is not an option. Instead, heuristics designed for graph problems give much better
performance. The GPS algorithm scales linearly with the number of edges
$\left|E\right|$, and since in a tight-binding representation $\left|E\right|
\propto N_\mathrm{grid}$, its algorithmic complexity is $\mathcal{O}(N_\mathrm{grid})$,
whereas the BFS partitioning algorithm scales as 
$\mathcal{O}(N_\mathrm{grid} \log N_\mathrm{grid})$. In any case, the scaling is much 
more favorable than that of the RGF algorithm, $\propto \sum_{i=0}^{N+1} M_i^3$,
so that for systems of typical size the overhead of the reordering process becomes
negligible, as we will demonstrate in the next section.

\subsection{Example: ring geometry}

In order to demonstrate the performance of the algorithms discussed above, 
we consider their application to a ring geometry in finite difference approximation.
The performance of the RGF algorithm after matrix reordering is compared with
the performance using the ordering that arises naturally in finite difference grids
as shown in Fig.~\ref{BTDform} (in the remainder referred to as \emph{natural partitioning}).

\begin{figure}
\includegraphics[width=\linewidth]{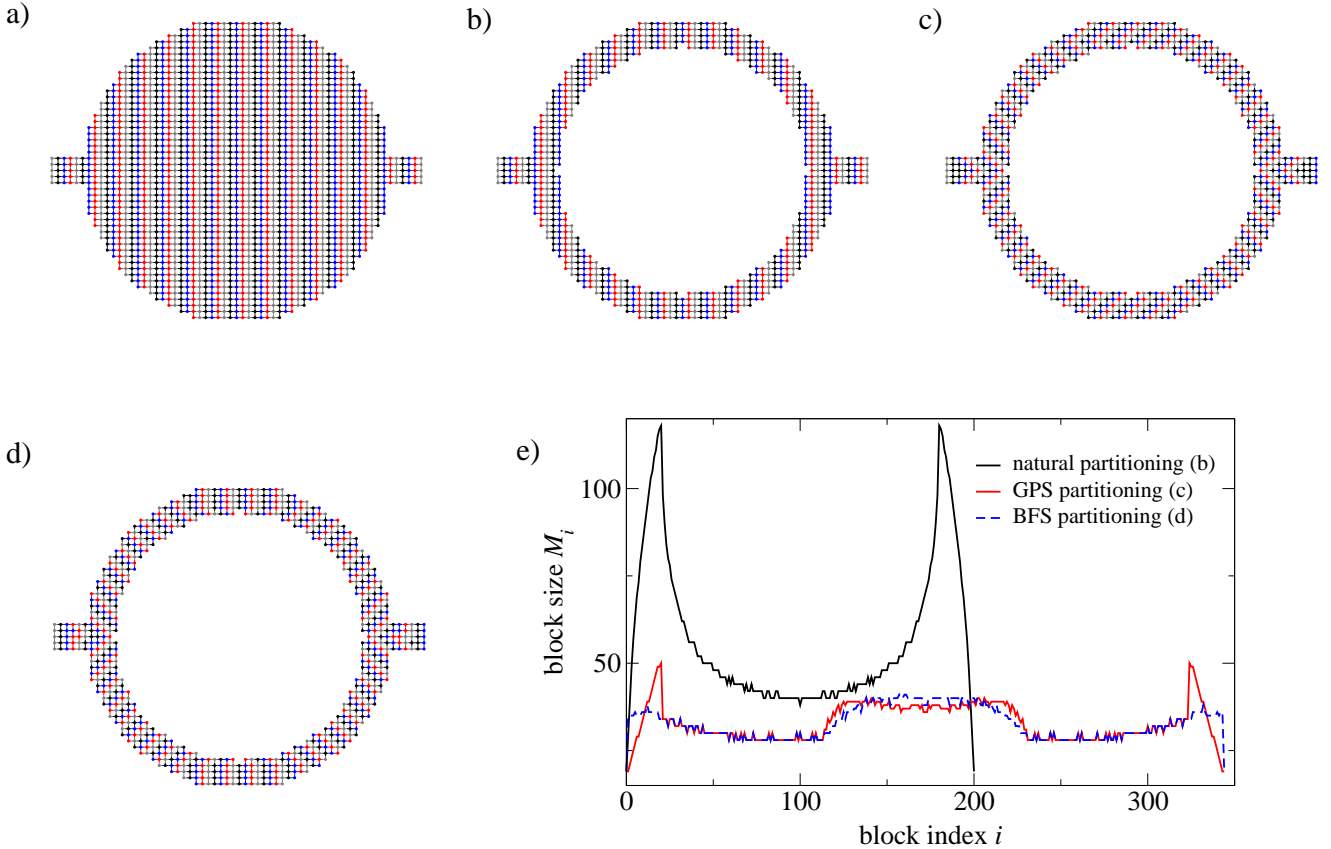}
\caption{Partitioning of the different systems considered in the
performance study. (a) Circular cavity in natural partitioning, 
where the ring structure is enforced by a potential term in the middle
of the cavity, (b) ring in natural partitioning, (c) ring 
in GPS partitioning and (d) ring in BFS partitioning. In (a)-(d), 
vertices belonging to the same level in the partitioning are marked
with the same color, different levels are marked in alternating colors. 
Note that, in order to reveal the partitioning structure more clearly, 
the grids shown here are much smaller than in a typical calculation.
(e) Size $M_i$ of the blocks $H_{ii}$ in the block-tridiagonal matrix for
a ring with 20 lattice points in the leads for
natural, GPS and BFS partitioning (the circular cavity is not shown here).
}\label{partitioning}
\end{figure}

In Fig.~\ref{partitioning}(a)--(d) we show four different approaches for
calculations in a ring geometry. A ring can be treated as a circular cavity
(see Fig.~\ref{partitioning}(a)), with a large potential on the lattice points inside the 
inner ring diameter. This approach is easier to implement than a real ring but
less efficient, as a large additional number of lattice points enters the
calculation. However, this approach has been used frequently in the past, and therefore
we also consider its performance. Transport calculations in a real ring
require somewhat more bookkeeping because of the non-trivial geometry, but
can be easily done describing the grid as a graph. For the circular cavity,
we only consider natural partitioning, as shown in Fig.~\ref{partitioning}(a), for 
the ring we consider natural partitioning (Fig.~\ref{partitioning}(b)), GPS
partitioning (Fig.~\ref{partitioning}(c)) and BFS partitioning 
(Fig.~\ref{partitioning}(d)). 

The partitionings in Fig.~\ref{partitioning}(c) and (d) are dramatically different from the
natural partitioning. The levels align mainly in vertical or diagonal
directions as these are the preferred directions in the square lattice. The number of
levels is increased with respect to the natural partitioning, as the distance between both
leads is much larger than the number of vertical slices, leading to a larger number of
blocks in the block-tridiagonal matrix, as can be seen from Fig.~\ref{partitioning}(e). 
Both GPS and BFS partitioning lead to a drastically reduced block size with respect to 
the natural partitioning and the result is rather balanced. Though
the actual partitionings in Fig.~\ref{partitioning}(c) and (d) look quite different, with
respect to minimizing the weight $w(H)$ they are equally good. The BFS partitioner 
conforms to the geometric structure, as it puts vertices in levels according to their distance 
from the leads. Except for a small number of blocks in the beginning and the end
of the block-tridiagonal matrix (see Fig.~\ref{partitioning}(e)), the GPS
partitioner leads to an equally balanced structure, although the partitioning looks quite 
different. The GPS partitioner works well in this case, as the leads are almost 
at maximum distance in this ring structure.

\begin{figure}
\begin{center}
\includegraphics[width=0.9\linewidth]{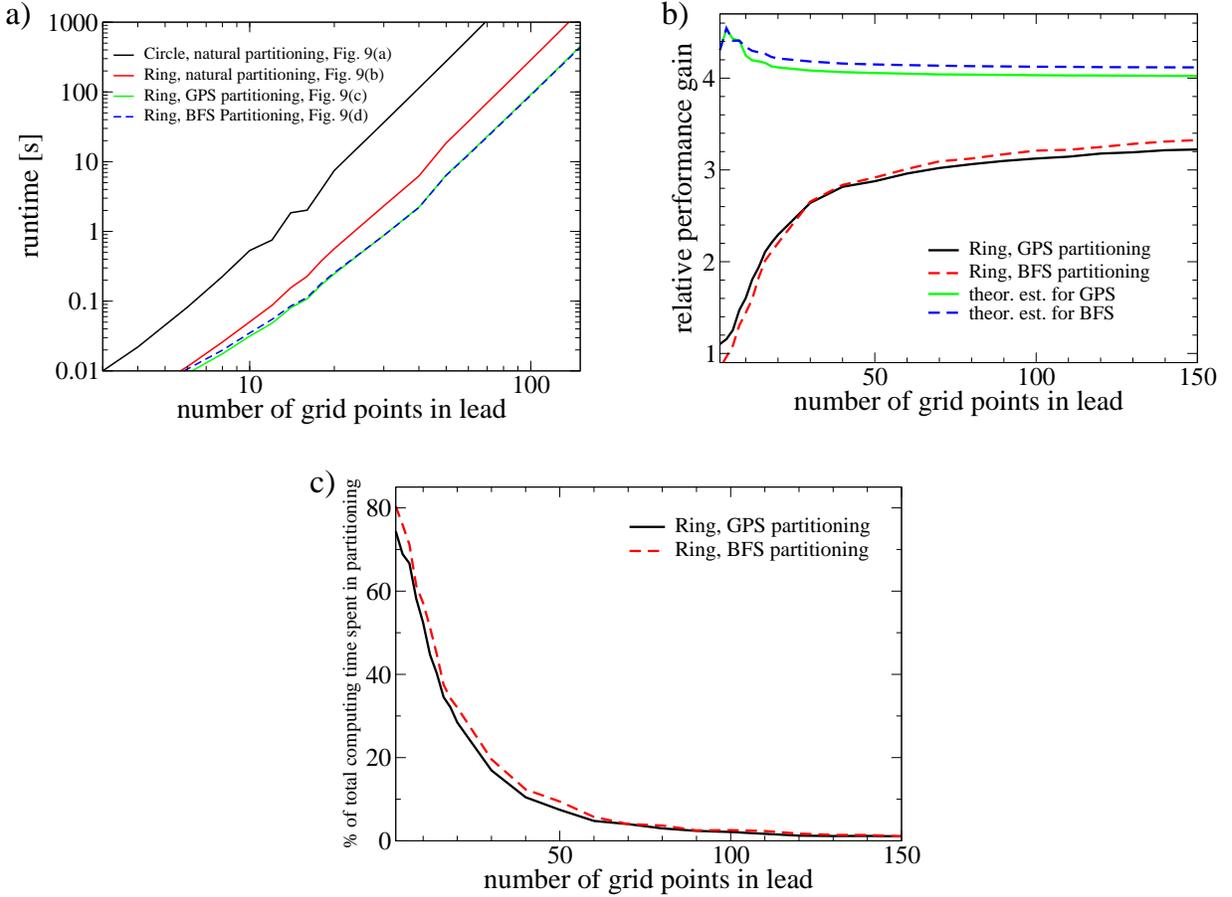}
\end{center}
\caption{Performance of the partitionings (for rings, Fig.~\ref{partitioning}): 
(a) runtime of the different
systems as a function of the number of lattice points in the leads, i.e. 
as a function of system size. The calculations were performed 
on a Core2Duo T5500 processor and 1GB of memory, and the runtime includes both partitioning
overhead and transport calculation. (b) Performance gain of GPS and BFS partitionings
with respect to natural partitioning and (c) overhead of the matrix reordering
as a function of system size.}\label{performance}
\end{figure}

We now apply the recursive Greens function algorithm from Sec.~\ref{algorithms} to
the different partitionings. In Fig.~\ref{performance}(a) we show the runtime
of the algorithm as a function of the number of lattice points in the leads, which is also the number
of lattice points across the arms of the ring. Note that in all cases the runtime includes
both the time spent in calculating the matrix reordering and the time spent in the
actual transport calculation.

The runtime scales similarly
in all cases, as this is the scaling of the RGF algorithm. Nevertheless, the runtimes of the different
approaches can differ by a factor that is significant.
As expected, the circular cavity is slowest, due to the extra number of lattice points. 
GPS and BFS partitionings lead to a rather similar performance that is \emph{significantly}
better than the ring in natural partitioning. It outperforms 
the circular cavity even by a factor of up to 100. In the remainder, 
we examine the performances of the ring for different partitionings in more detail and
leave out the circular cavity. In Fig.~\ref{performance}(b) we show the 
relative performance gain of GPS and BFS partitionings over the natural
partitioning. Except for the smallest of systems that are too small
to be useful in practice, the performance of
GPS and BFS partitionings is better than the natural partitionings. Even for moderately
sized systems the performance gain through the matrix reordering is 
approximately~3, with the BFS
partitioner being slightly better than the GPS partitioner.
In Fig.~\ref{performance}(b) we also show estimates of the performance gain calculated 
from the weights $w(H)$ of the different partitionings. These estimates predict
a performance gain of approximately~4. For small system sizes, we see deviations
from these estimates because of the overhead of the partitioning process, for larger
system sizes we almost reach the estimated value. On modern
computer architectures, runtime does not only depend on the number of arithmetic
operations~\cite{ATLAS}, and thus we do not achieve the full theoretical
potential of the reordering, yet still significant improvements. 
In Fig.~\ref{performance}(c) we show the fraction of time spent in calculating the matrix
reordering with respect to the total computation time, and as expected the overhead becomes
negligible already for moderately sized systems. It should be noted that in actual
calculations the partitioning is commonly
only done once, and transport calculations can be 
done repeatedly with the same partitioning: Usually one is interested
in transport properties depending on some parameters, and these generally 
do not change the \emph{structure} of the Hamiltonian matrix but only the 
values of the respective entries. In this case, the partitioning overhead becomes even
more irrelevant.  

\begin{figure}
\begin{center}
\includegraphics[width=0.4\linewidth]{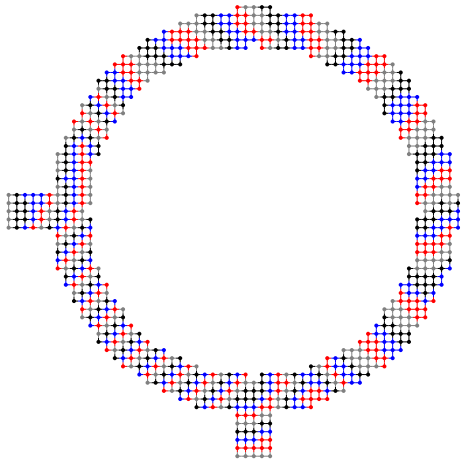}
\end{center}
\caption{BFS partitioning of a ring with perpendicular leads. Note that in this case
the two leads are closer together and thus the number of blocks is reduced. Therefore
the block sizes tend to be larger than in the previous examples.}\label{complexstructure}
\end{figure}

It should be emphasized, that for all the situations, the \emph{same} 
transport code was used. In addition to the significant speedup through the 
graph techniques considered here, the abstraction of the system through graph structures
allows for very generic transport codes. This is an additional strength 
of this approach, as the well-established RGF algorithm is thus readily applied 
to arbitrary systems that would require special treatment otherwise, such as a
scattering region with perpendicular leads, as depicted in Fig.~\ref{complexstructure}.

Of course, for any system, an algorithm designed for that special system will probably
outperform generic approaches. However, the combination of matrix reordering and 
RGF algorithm can be applied to arbitrary systems and is thus probably the most
versatile transport approach. In addition, all the algorithms
relying on the block-tridiagonal structure of the Hamiltonian mentioned
in Sec.~\ref{algorithms} can benefit from these matrix reordering strategies. 

Modern transport calculations tend to be more and more complex and time-consuming. For example,
an electronic calculation including the electron spin typically takes $2^3=8$ times longer
than a calculation on spin-less electrons. Furthermore, calculations including disorder
involve averages over many disorder configurations. Any increase in computation speed 
is therefore beneficial, and the added versatility through matrix
reordering methods makes these techniques even more useful.

\section{Spin filtering in nanostructures}\label{spinfilter}
In the introductory example in Sec.~\ref{sectionintroduction} it was shown 
how to realize systems that work as spin switches making use of the interference of 
wavefunctions propagating clockwise and counterclockwise in Aharonov-Bohm rings
with SO-interaction.
However, several other device proposals have been put forward making use of 
different concepts in order to achieve spin filtering in mesoscopic 
systems.

A very prominent category is transverse focusing of 
ballistic electrons/holes in two dimensional electron/hole gases (2DEGs/2DHGs)~\cite{Houten1989}. 
In materials that exhibit SO interaction the cyclotron radius of ballistic 
electrons/holes due to a magnetic field perpendicular to the 2DEG/2DHG  depends 
on the spin state of the charge carriers. Therefore, it is possible to filter out 
either spin-up or spin-down electrons by an appropriate arrangement of quantum point 
contacts and a proper choice of the perpendicular magnetic field. Apart from 
several theoretical treatments of this topic~\cite{Usaj2004,Govorov2004,Reynoso2007}, 
spin filtering by transverse focusing has already been experimentally verified in a GaAs-based 
2DHG.~\cite{Rokhinson2004} In related experiments, spin-polarized currents in 2DEGs/2DHGs were 
detected by a setup consisting of point contacts and making use of transverse 
focusing of the charge carriers~\cite{Potok2002}. With such a detector it was possible to 
confirm the presence of spin-polarized currents emitted 
from mesoscopic quantum dots utilizing quantum interference at high in-plane 
magnetic fields~\cite{Folk2003}  and
from quantum point contacts, which were either made spin sensitive with high 
in-plane magnetic 
fields~\cite{Potok2002} or showed a pronounced "0.7-anomaly"~\cite{Rokhinson2006}.

A further appealing approach to filter spins 
are three terminal structures, that act as mesoscopic Stern-Gerlach type spin 
filters~\cite{Fabian2002}. In these devices one of 
the leads injects spin-unpolarized current and, after passing a region where 
the spin-degeneracy is lifted, oppositely-polarized output currents exit 
through the other two leads. 
This separation of up and down spins can be accomplished, e.g., by utilizing 
Rashba SO interaction~\cite{Kiselev2001,Ohe2005,Cummings2006}.

However, three terminals are not required to create spin 
polarized currents. Many devices, as e.g. the AB-ring presented in
Sec.~\ref{sectionintroduction},
typically rely on two terminals only, where transport through tailored geometries with SO 
interaction~\cite{Eto2005,Zhai2007,Scheid2007}, 
magnetic fields~\cite{Song2003,Zhai2006,Scheid2007a} 
or a combination of both~\cite{Zhai2005} 
can result in a significant spin filter effect.

As a representative example for the methods mentioned above,
in the present section we present spin filtering 
due to Rashba SO interaction in quantum wires 
connected to two terminals. 
We consider a quantum wire in $y$-direction realized in a 2DEG in the 
($x,y$) plane connected to two nonmagnetic leads. The
Hamiltonian of the system, whit spatially dependent Rashba SO interaction 
is given by
%
%
\begin{equation}\label{hamiltonian}
H_0 = \frac{\hat{p}^2}{2m^*} + 
\frac{\alpha (x)}{2\hbar} (\hat{\sigma}_x \hat{p}_y -
\hat{\sigma}_y \hat{p}_x) +
(\hat{\sigma}_x \hat{p}_y -
\hat{\sigma}_y \hat{p}_x) \frac{\alpha (x)}{2\hbar} 
+ V(x,y) + U_\text{B}(x,y) \, .
\end{equation}
%
%
Here $V(x,y)$ is the lateral transverse confinement potential forming the quantum wire, 
while $U_\text{B}(x,y)$ is an additional electrostatic potential in the system, e.g., 
realized by gate voltages. Furthermore, $\hat{\sigma}_i$ denote the Pauli spin matrices,
and $m^*$ is the effective electron mass of the semiconducting material.
We consider a constant Rashba SO interaction strength 
$(=\alpha _\text{C})$ in the central region of the system, which is connected to 
two semi-infinite leads on opposite sides, where $\alpha (x)$ is chosen to be zero avoiding 
ambiguities in the definition of spin current that arise for leads with SO 
interaction~\cite{Shi2006}. 
In order not to 
introduce additional effects due to an abrupt jump in the SO coupling strength, 
the parameter $\alpha (x)$ is changed sufficiently smooth from zero to $\alpha _\text{C}$ 
between the leads and the central region.
For the numerical calculations presented in the next two sections 
the Hamiltonian~(\ref{hamiltonian}) is discretized 
as shown in Sec.~\ref{sectiontransport} yielding a tight-binding Hamiltonian 
on a square grid.

In the rest of this section we investigate the transport properties in the linear response
regime due to an infinitesimal bias voltage $\delta U$ applied between the left and right 
contact. The charge (C) and spin (S) current in the 
Landauer formalism are then given by
\begin{displaymath}
I_\text{C/S}=G_\text{C/S}T_\text{C/S}\delta U,
\end{displaymath}
where $G_{\text{C}}=e^2/h$ and $G_{\text{S}}=e/4\pi$ are the conductance quanta 
of charge and spin respectively. 
Since, opposite to charge current, the spin current can be different in the right 
and left lead~\cite{Scheid2007a}, here we choose to evaluate the respective currents 
in the right lead. Then the transmission probabilities $T_\text{C/S}$ at the 
Fermi energy are 
given by
\begin{equation}
\label{transmissions}
T_\text{C}= T_{+,+}+T_{+,-}+T_{-,+}+T_{-,-},\qquad
T_\text{S}= T_{+,+}+T_{+,-}-T_{-,+}-T_{-,-},
\end{equation}
where $T_{\sigma,\sigma '}  \! = \! \sum_{ n \in \text{R},n' \in \text{L}} 
\left| S_{n,\sigma;n',\sigma'}\right|^2$ is the probability for an electron, 
injected into the left (L) lead with spin state $\sigma '$ to be transmitted to 
the right (R) lead and end up there in spin state $\sigma$. 
In the present and the following section we fix the spin quantization axis 
to the $y$-axis. The scattering matrix elements $S_{n,\sigma;n',\sigma'}$ 
and therefore also the spin resolved transmission probabilities 
$T_{\sigma ,\sigma '}$ are evaluated using the recursive Greens function 
algorithm presented in Secs.~\ref{sectiontransport} and~\ref{sectionreordering}. 

One general feature of Landauer transport in a quantum wire with SO interaction 
and non-magnetic leads is the absence of spin-polarized currents 
in a lead that supports only a single transversal mode. This property can be 
derived from the invariance of the system under the time-reversal operator 
$\hat{\mathcal{T}}=-\mathrm{i}\hat{\mathcal{C}}\sigma _y$~\cite{Zhai2005a}, 
where $\hat{\mathcal{C}}$ is the operator of complex conjugation.
For a perfect quantum wire which is translationally invariant 
in the direction of transport all occupied transversal subbands transmit 
without reflection and spin polarization is not possible due to 
SO interaction. However, if backreflection, caused by deviations from a perfect 
quantum wire, is present, it is possible to observe spin-polarized currents 
in leads with at least two transversal channels.
There the typical mechanism responsible for spin 
polarization is the mixing of spins from different transversal subbands due to 
the SO interaction.\\
In Eq.~(\ref{hamiltonian}) this translational invariance in $y$-direction is already broken 
by the spatially varying SO interaction $\alpha (x)$
even if the quantum wire was perfect in $y$-direction otherwise, i.e. $V(x,y)=V(y)$ and 
$U_\text{B}(x,y)=U_\text{B}(y)$. However, if the region where $\alpha (x)$ is turned on/off 
is sufficiently long, reflection due to the change of $\alpha (x)$ is negligible.\\
There exist several device proposals relying on this mixing of spins from different 
transversal subbands due to $x$-dependent lateral confinement potentials $V(x,y)$ 
or other superimposed electrostatic potentials $U_\text{B}(x,y)$. These device 
designs include, e.g., constrictions~\cite{Eto2005,Silvestrov2006}, 
lateral side pockets~\cite{Zhai2007}, 
or electrostatic barriers~\cite{Scheid2007}, to name only a few.\\
In most of those proposals, systems symmetric with respect to inversion of 
the $x$-coordinate were considered, i.e. 
$V(x,y)=V(-x,y)$, $U_\text{B}(x,y)=U_\text{B}(-x,y)$ and 
$\alpha (x)=\alpha (-x)$. Then 
the Hamiltonian~(\ref{hamiltonian}) is left invariant upon application of 
the symmetry operation
\begin{equation}\label{sym:linres}
\hat{\mathcal{P}}=-\mathrm{i}\hat{\mathcal{C}}\hat{R}_x\hat{\sigma} _z,
\end{equation}
where the operator $\hat{R}_x$ inverses the $x$-coordinate.
The scattering wavefunctions inside the leads are changed by the operator 
$\hat{\mathcal{P}}$ in the following way:
$\hat{R}_x$ exchanges the leads, i.e., a transversal mode index $n$ 
is replaced by the corresponding mode index $\bar{n}$ in the other lead. 
The operator of complex conjugation transforms incoming (outgoing) states 
into outgoing (incoming) states with complex conjugated amplitude.
Moreover, the combined effect of $\hat{\mathcal{C}}\sigma _z$ is to rotate 
a spinor with the coordinates $(\theta , \phi)$ on the Bloch sphere 
to the coordinates $(\theta , -\phi +\pi)$. Exploiting this invariance 
of the Hamiltonian one can derive the relation 
$S_{n,\sigma;n',\sigma '}(E)=S_{\bar{n}',\sigma ';\bar{n},\sigma}(E)$
between the elements of the scattering matrix (see
Refs.~\cite{Zhai2005a,Scheid2007a} for related expressions).
This results in the equality of the spin flip transmissions $T_{+,-}=T_{-,+}$. 
Therefore, for those devices to be able to work as a spin filter, in view of  
Eq.~(\ref{transmissions}) the spin conserving 
transmissions $T_{+,+}$ and $T_{-,-}$ need to be different.\\
%
%
\begin{figure*}
	\begin{center}
	\includegraphics[width=0.9\textwidth]{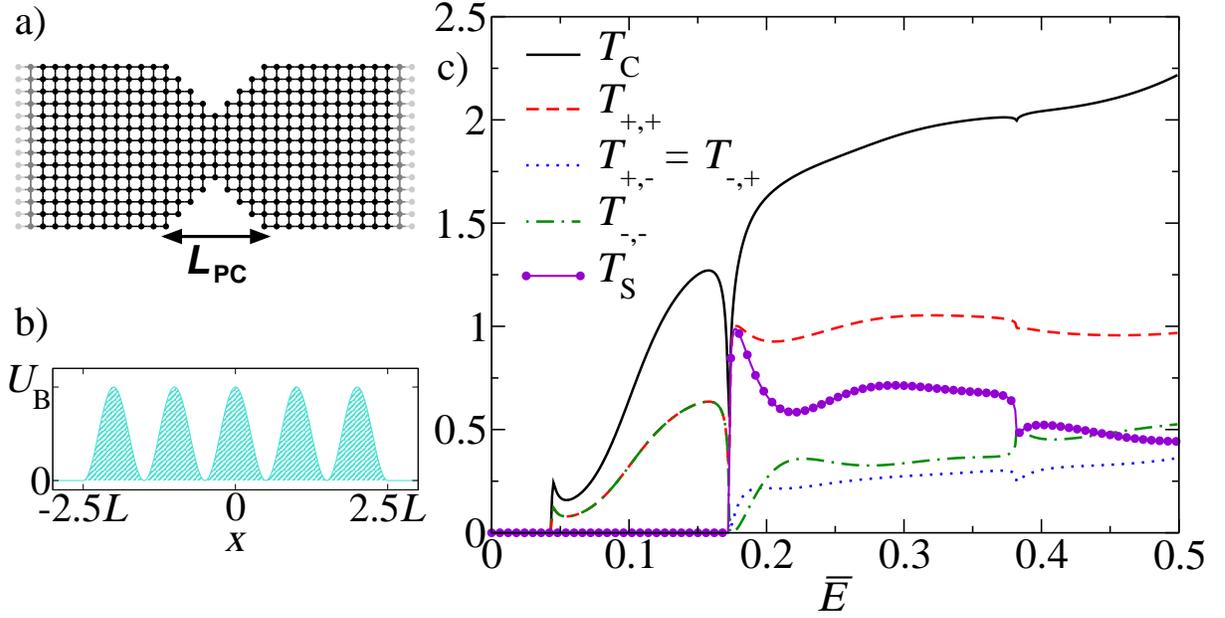}
	\caption{\label{fig:one} Panel a) The square lattice discretization 
	of a quantum wire (width $W= 15a$) with a single constriction of length 
	$L_\text{PC}=10a$, see also Eq.~(\ref{constriction}).
	Panel b) Periodic array of $N=5$ electrostatic barriers with period length $L$ and 
	barrier height $U_\text{B}$. 
	Panel c) Charge $T_\text{C}$, spin $T_\text{S}$ and 
	spin resolved transmission probabilities $T_{\sigma ,\sigma '}$ for the system depicted in panel a) and 
	specified in the text at fixed SO interaction strength 
	$\bar{\alpha}=\frac{m^*a}{\hbar ^2}\alpha _\text{C}=0.1$ with respect to the Fermi energy 
	$\bar{E}=\frac{2m^*a^2}{\hbar ^2}E$. The $n$-th transversal mode in the wire opens at
	$\bar{E}_n=\frac{\pi ^2n^2}{(W/a)^2}.$}
	\end{center}
    \end{figure*}
%
%
As an example, in the following we consider Landauer transport 
through a Rashba SO quantum wire 
with a constriction. Similar calculations were 
carried out in Refs.~\cite{Eto2005,Reynoso2007} where it was shown, that 
this setup is able to produce a spin polarized current of sizeable quantity. 
It was conjectured, that the mechanism responsible for the spin polarization 
was the depletion of higher transversal modes of the wire and a subsequent 
spin dependent repopulation of those modes when traversing the constriction~\cite{Eto2005}.
To experimentally observe the predicted spin polarization the use of a transverse electron 
focusing technique was suggested~\cite{Reynoso2007}.

\begin{figure*}
	\begin{center}
	\includegraphics[width=0.6\textwidth]{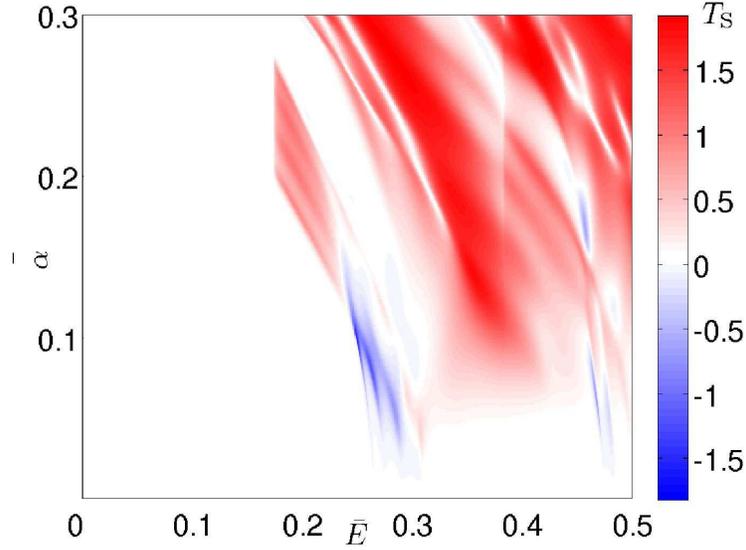}
	\caption{\label{fig:three} Spin transmission probability $T_\text{S}$ plotted 
	as a function  of the Fermi energy 
	$\bar{E}$ and SO interaction strength $\bar{\alpha}$ for a quantum wire 
	with $N=5$ electrostatic barriers of length $L=10a$ and height 
	$\bar{U}_\text{Barr}=\frac{2m^*a^2}{\hbar ^2}U_\text{Barr}=0.2$. 
	}
	\end{center}
    \end{figure*}

A typical grid (with lattice spacing $a$) used in the calculation for the symmetric point 
contact in a wire of width  $W=15a$ is shown in Fig.~\ref{fig:one}a, 
where the constriction of length $L_\text{PC}=10a$ is formed by hard-wall potentials: 
\begin{equation}\label{constriction}
V(x,y)=\begin{cases}0 &\mathrm{for}\;C(x)<y<15a-C(x)\\\infty &\mathrm{else}
\end{cases}\quad\;\mathrm{with}\;
C(x)=\begin{cases}2.05a \left( 1-\cos \left( \frac{2\pi
(x+L_\text{PC}/2)}{L_\text{PC}}\right)\right)&\mathrm{for}\;|x|<L_\text{PC}/2\\0
&\mathrm{otherwise.}
\end{cases}  
\end{equation}
Additionally $U_\text{B}(x,y)$ is set to zero. 
In Fig.~\ref{fig:one}c we present the relevant transmission probabilities 
with respect to the Fermi energy $E$ for this system. 
There we observe that the total conductance $T_\text{C}$ is reduced in comparison with 
that of a perfect quantum wire. In the latter case 
$T_\text{C}$ exhibits sharp steps due to conductance quantization~\cite{Wees1988,Wharam1988} 
which are washed 
out here due to tunneling processes through the constriction. 
Furthermore, for energies, where only a single transversal mode is 
supported in the quantum wire, the spin transmission vanishes as expected, i.e. 
$T_{+,+}=T_{-,-}$ for energies $\bar{E}\lesssim0.175$. 
Also the relation 
$T_{+,-}=T_{-,+}$ is fulfilled as required by the symmetry of the setup. 
Finally, at energies where a new transversal mode opens up 
($\bar{E}_2\approx0.18,\;\bar{E}_3\approx0.39$) 
dips in the transmission probabilities become apparent. Those dips can be 
explained by interference of localized states in the central region where 
$\alpha(x)=\alpha _\text{C}$ and the scattering states in the quantum
wire where $\alpha(x)=0$~\cite{Sanchez2006}.\\
%
%
\begin{figure*}
	\begin{center}
	\includegraphics[width=0.6\textwidth]{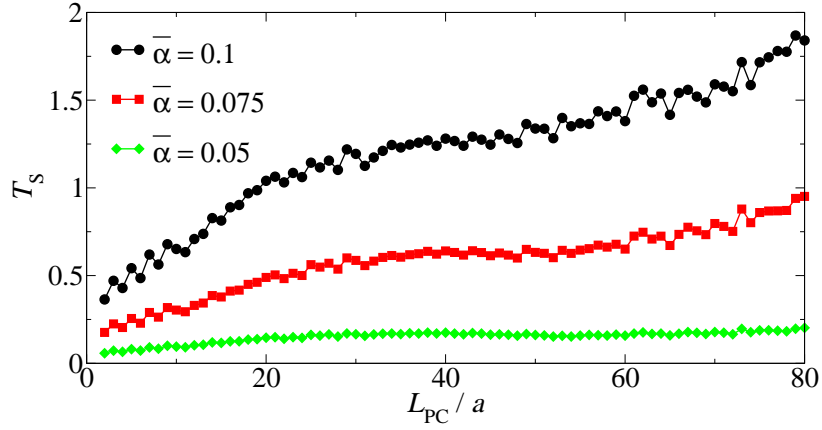}
	\caption{\label{fig:two} Spin transmission probability $T_\text{S}$ at fixed injection energy 
	$\bar{E}=0.25$ within the second transversal mode for three different SO interaction 
	strengths $\bar{\alpha}=0.1$ (black dots), $\bar{\alpha}=0.075$ (red squares) and 
	$\bar{\alpha}=0.05$ (green diamonds) with respect to the length of the constriction $L_\text{PC}$.}
	\end{center}
    \end{figure*}
%
%
In order to study the influence of the adiabaticity of the constriction on the degree of 
spin polarization that can be extracted, in Fig.~\ref{fig:two} we show $T_\text{S}$
as a function of the length of the constriction $L_\text{PC}$ for 
several values of $\alpha _\text{C}$. For all of the curves we clearly observe an increase 
in spin transmission 
with increasing $L_\text{PC}$, in accordance with the mechanism suggested in Ref.~\cite{Eto2005}. 
There it was argued in the context of Landau-Zener transitions 
that the repopulation of higher transversal subbands will be more efficient 
for more adiabatic constrictions or barriers, resulting in a higher 
degree of spin polarization. For $\bar{\alpha}=\frac{m^*a}{\hbar ^2}\alpha _\text{C}=0.1$ 
the spin transmission even approaches the maximal possible value $T_\text{S}=2$.
One drawback of the presented system is its restriction 
to unidirectional spin polarization. In agreement with the model of Ref.~\cite{Eto2005}, 
Figs.~\ref{fig:one} and~\ref{fig:two} give evidence that $T_\text{S}\geq 0$ for the parameter range 
considered. This limitation to output current with fixed spin polarization direction 
restricts the usability of the spin filter to special purposes. A possible way to 
circumvent this constraint is the use of a periodic array of electrostatic 
barriers~\cite{Scheid2007}, which we now briefly investigate. 
Fig.~\ref{fig:three} shows the spin transmission 
of a straight hard-wall quantum wire ($C(x)=0$ in  Eq.~(\ref{constriction}))
subject to $N=5$ electrostatic barriers,
\begin{displaymath}
U_\text{B}(x,y)=U_\text{B}(x)=\begin{cases}U_\text{Barr}\left( 0.5-0.5\cos 
\left(\frac{2\pi(x+L/2)}{L}\right)\right)&\mathrm{for}\;-NL/2<x<NL/2\\0
&\mathrm{otherwise} \, , 
\end{cases}
\end{displaymath} (as shown in Fig.~\ref{fig:one}b). The spin transmission is plotted
as a function of the Fermi energy and the SO interaction strength. 
Again, one observes $T_\text{S}= 0$ for energies below $\bar{E}_2$. Furthermore, 
different regions in parameter space exhibit opposite sign of $T_\text{S}$,
enabling to change the sign of the output polarization, e.g.,
by tuning $\alpha _c$ via gate voltages~\cite{Nitta1997}. This additional functionality 
is due to resonant tunneling, which is absent for a quantum wire with only 
a single constriction or barrier.

\section{Pure spin current generation}\label{purespin}

In the preceding sections we focused on mesoscopic geometries that exhibit 
functionalities such as spin filtering or spin switching when applying an external bias
between the contacts in the system. In the case of a static dc bias in a two terminal 
geometry those spin currents arise due to different currents of spin-up and 
spin-down electrons flowing into the direction of the contact with the lower chemical 
potential. However, over the last few years a growing number of device proposals has 
been put forward, that exhibit the interesting feature of pure spin current generation,
i.e. spin currents in the absence of net charge transport. This intriguing case 
appears, when the direction of motion of spin-up electrons is opposite to the direction 
of motion of spin-down electrons and both currents equal in absolute magnitude.

Among the devices sharing the prospect of creating pure mesoscopic spin currents 
are systems with more than two terminals realized in 2DEGs 
exhibiting  Rashba~\cite{Rashba1960,Bychkov1984} and/or the
Dresselhaus~\cite{Dresselhaus1955} SO coupling.
Here a charge current is induced in this multiterminal structure by the application of 
bias voltages between the different contacts of the system. 
However, if the voltage of one of the leads is adjusted to make it work as a voltage probe, 
no charge current passes this lead but a pure spin current can appear owing to the SO 
coupling present in the system. The basic working principle behind these devices is the 
so-called mesoscopic spin Hall effect~\cite{Bardarson2007}, 
a version of the intrinsic spin Hall effect 
\cite{Sinova2004}, where the typical system size does not exceed the phase 
coherence length of the electrons. The SO interaction leads to different transport 
dynamics for different spin species, which can be used to extract the desired pure spin 
currents by a clever design of the multiterminal 
geometry~\cite{Hankiewicz2004,Sheng2005,Souma2005}.\\
Complementary to the generation of pure spin current in multiterminal geometries, there are 
other types of devices not relying on the application of a net dc-bias. 
In spin pumping, the cyclic variation of two or more system parameters, such as e.g. gate 
voltages, induces spin-polarized currents at zero bias, where the induced charge current 
can be tuned to disappear, leaving pure spin currents. Several realizations of spin pumps 
in mesoscopic systems have been proposed, relying on SO
interaction~\cite{Governale2003,Sharma2003} or the Zeeman coupling of 
electrons to external magnetic fields~\cite{Mucciolo2002}. The latter proposal has
been experimentally confirmed~\cite{Watson2003} by detecting spin-polarized currents 
making use of a transverse electron focusing technique~\cite{Potok2002}
mentioned in the previous section. 

Complementary to pumps, ratchets only require a single driving parameter to 
achieve directed transport, and the current direction can be switched upon
tuning external parameters such as temperature. In addition to the requirement of a 
broken spatial symmetry the ratchet has to be operated out of thermal equilibrium.
The concept of particle ratchets, which has been addressed in numerous
works~\cite{Reimann2002}, 
has recently been extended to systems called spin ratchets. To be specific, the mesoscopic 
spin ratchets proposed so far~\cite{Scheid2006,Scheid2007,Scheid2007a}, 
are based on a quantum wire realized in a 2DEG.  Between the two contacts attached to the 
quantum wire an ac bias voltage $U_\text{R}(t)$ is applied (rocking ratchet)
with zero net (time-averaged) bias. Furthermore, in the central region of the quantum wire 
the spin degeneracy is lifted due to either SO interaction~\cite{Scheid2007} 
or the Zeeman coupling to an external magnetic field~\cite{Scheid2006,Scheid2007a}. 
Upon appropriate choice of the 
system geometry and tuning of the external driving the charge transported in the forward 
($U_\text{R}>0$) and backward bias ($U_\text{R}<0$) situation can be made equal
allowing for spin currents in the absence of net charge transport.
\\
In the following we outline the model for the spin ratchets introduced in 
Refs.~\cite{Scheid2006,Scheid2007,Scheid2007a}. There, driving with a period $t_0$ is considered. 
It is implied, that this period is 
much larger than characteristic time scales related to the electron transport through the quantum 
wire. Therefore, the system is assumed to be in a steady state at every 
instance of time, and the Landauer-B\"uttiker approach to transport is used 
for the computation of the ratchet currents. To be specific, 
we consider an unbiased square wave driving 
$U_\text{R}(t)=U_0 \,\text{sign} \left[ \sin (2\pi t/t_0) \right]$, where 
$U_\text{R}(t)$ is restricted to the values $\pm U_0$. 
The net current is then given by the average of the 
steady-state currents in the two rocking situations 
(labeled $+U_0$ and $-U_0$ respectively) for both charge and spin, 
\begin{equation} 
\langle I_\text{C/S}(U_0) \rangle = [I_\text{C/S}(+U_0)+I_\text{C/S}(-U_0)]/2.
\end{equation}
Since the spin ratchet effect requires nonlinear transport \cite{Scheid2006}, 
i.e.\ finite bias voltages,
the Hamiltonian~(\ref{hamiltonian}) introduced in the previous section has to be 
extended to additionally include the effective electrostatic potential in the 
conductor due to the applied bias. Therefore, we add the term
$H_\text{R}=eU_\text{R}g(x,y;U_\text{R})$ to the Hamiltonian~(\ref{hamiltonian}) 
yielding the full Hamiltonian at finite bias:
\begin{equation}\label{fullhamiltonian}
H=H_0+H_\text{R}.
\end{equation}
Furthermore at finite bias 
a generalized version of the expressions for charge and spin current valid at $U_\text{R}\neq 0$ 
has to be used. For coherent Landauer transport those currents can be obtained from an 
integration of the transmission probabilities over the Fermi window~\cite{Datta2002}. 
Finally, the averaged charge $\langle
I_\text{C} \rangle$ and spin $\langle I_\text{S} \rangle$ currents can
be written as~\cite{Scheid2007a}
%
%
\begin{displaymath}\label{current}
 \langle I_{\text{C/S}}(U_0) \rangle = \frac{G_{\text{C/S}}}{2e} \!
\int_{E_\text{C}}^{\infty} \!\! \text{d}E \, 
\Delta f(E,U_0)\Delta T_{\text{C/S}}(E,U_0) \, .
\end{displaymath}
%
%
Here $E_\text{C}$ is the energy of the 
conduction band edge and $\Delta f(E,U_0)= \left[ f(E,E_\text{F}+eU_0/2) -
f(E,E_\text{F}-eU_0/2) \right]$ is the difference between the 
Fermi functions of the leads at bias voltage $U_0$, defining the Fermi window.
The averaged charge/spin transmission is just the difference between the 
steady state transmissions in the two rocking situations:
%
%
\begin{equation}\label{ratchet:transmission}
\Delta T_{\text{C/S}}(E, U_0)=T_{\text{C/S}}(E,+U_0) -
T_{\text{C/S}}(E,-U_0).
\end{equation} 
%
%

Considering the Hamiltonian~(\ref{fullhamiltonian}) we now show 
under what conditions the net charge transported after one full rocking 
period is zero, i.e., $ \langle I_{\text{C}}(U_0)\rangle =0$. If the electrostatic 
potentials $V(x,y)$, $U_\text{B}(x,y)$ 
and the Rashba SO strength $\alpha (x)$ are 
invariant under inversion of the $x$-coordinate,
\begin{equation}\label{ratsym}
V(x,y)=V(-x,y),\qquad U_\text{B}(x,y)=U_\text{B}(-x,y),\qquad\alpha (x)=\alpha (-x) \, ,
\end{equation}
it is appropriate
to assume that the electrostatic potential distribution due to the finite applied 
voltage $g(x,y;U_\text{R})$ also possesses this symmetry. 
Then the total Hamiltonian~(\ref{fullhamiltonian}) is invariant under the action 
of the symmetry operation 
$\hat{\mathcal{P}}=-\mathrm{i}\hat{\mathcal{C}}\hat{R}_\text{U}\hat{R}_x\sigma _z$ where $R_\text{U}$ switches 
the sign of the bias voltage ($\pm U_0\leftrightarrow \mp U_0$), 
yielding the relation 
\begin{displaymath}
T_{\sigma ,\sigma '}(E,\pm U_0)=T_{\sigma ', \sigma}(E,\mp U_0)
\end{displaymath}
between the spin-resolved transmission probabilities 
in the two rocking situations~\cite{Scheid2007a}.
Inserting this relation into Eq.~(\ref{ratchet:transmission}), 
we observe that the expression for the net 
charge transmission $\Delta T_\text{C}$ is zero, resulting in vanishing net charge current. 
Furthermore it can be used to simplify the expression for the
net spin current:
\begin{displaymath}
\langle I_{\mathrm{S}}(U_0)\rangle =\frac{G_\text{S}}{e}\int_{E_\text{C}}^{\infty} \mathrm{d}E \;\Delta f(E;U_0)\Big[T_{+,-}(E,+U_0)-T_{-,+}(E,+U_0)\Big] \,.
\end{displaymath}
At $U_0=0$ the Hamiltonian~(\ref{fullhamiltonian}) reduces to Eq.~(\ref{hamiltonian}), 
which is invariant 
under the operation of Eq.~(\ref{sym:linres}), yielding $T_{+,-}(E,0)=T_{-,+}(E,0)$. 
This absence of net spin current $\langle I_{\mathrm{S}}(U_0=0)\rangle=0$ in the linear response 
regime is in agreement with the theoretical prediction~\cite{Scheid2007a}. However, for finite rocking 
voltages $U_0\neq 0$ the additional potential introduced via $H_\text{R}$ breaks this 
symmetry and therefore enables different spin flip transmissions and thus 
a resulting net spin current.

We now turn our attention to the quantum wire 
 shown in Fig.~\ref{fig:one}a. For this system, which exhibits the 
symmetries of Eq.~(\ref{ratsym}), we perform numerical calculations at finite bias $U_0$, 
in order to confirm its operability as a spin ratchet. 
%
%
\begin{figure*}
	\begin{center}
	\includegraphics[width=0.6\textwidth]{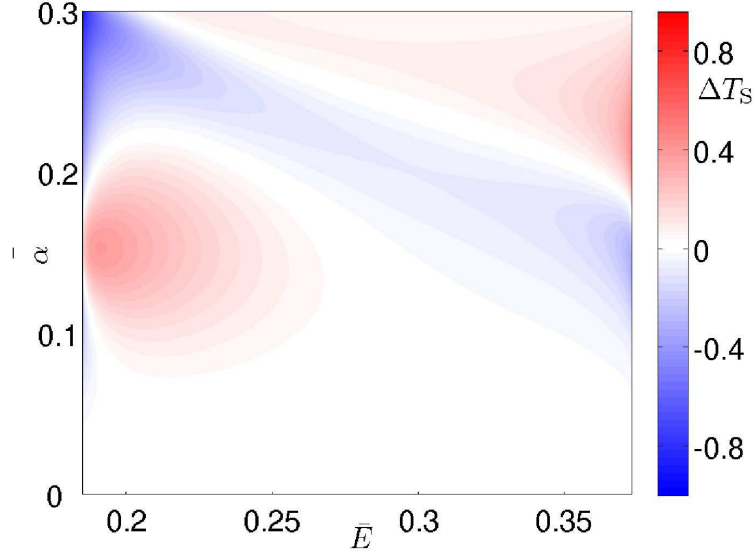}
	\caption{\label{fig:four} Net ratchet spin transmission probability 
	$\Delta T_\text{S}(U_0)=2\left( T_{+,-}(+U_0)-T_{-,+}(+U_0)\right)$ 
	for a stripe with a junction (see text) presented as a function of
	injection energy 
	$\bar{E}$ and SO interaction strength $\bar{\alpha}$ at finite applied voltage 
	$eU_0=0.02\frac{\hbar ^2}{2m^*a^2}$. Note the sign change in the 
	spin transmission upon tuning the SO coupling. }
	\end{center}
    \end{figure*}
%
%
In general, the function $g(x,y;U_\text{R})$ has to be obtained by self-consistently 
solving the Schr\"odinger and Poisson equation of the system. 
However, in the present treatment we make use of a 
simple model for $g(x,y;U_\text{R})$ assuming a linear voltage drop in the region where the 
point contact is formed in the quantum wire: 
\begin{displaymath}
g(x,y;U_\text{R})=g(x)=\begin{cases}1/2 & \mathrm{for}\;x<-L_\text{PC}/2 \, , \\
-x/L_\text{PC} & \mathrm{for}\;-L_\text{PC}/2<x<L_\text{PC}/2 \, , \\
-1/2 & \mathrm{for}\;x>L_\text{PC}/2 \, .
\end{cases}
\end{displaymath}
It is well known that a constriction in a quantum wire acts as an effective potential barrier 
constituting a region where the voltage applied across the wire is likely to drop. 
Since the bias voltages we consider are small compared to the energy shift introduced by the constriction, 
this assumption of a linear voltage drop should be an appropriate approximation
for the actual distribution of the 
electrostatic potential in this wire~\cite{McLennan1991}.

In Fig.~\ref{fig:four} we plot the net spin transmission $\Delta T_\text{S}$ as a
function of 
$\bar{\alpha}$ and the injection energy in the range, where both leads support two transversal 
modes. We observe that $|\Delta T_\text{S}|$ reaches values of up to 0.9 in the parameter 
range shown. Furthermore, for a given value of injection energy, the sign of $\Delta T_\text{S}$ can be 
switched by changing $\alpha _\text{C}$. Since the strength of the Rashba 
SO interaction $\alpha _\text{C}$ can be tuned via gate 
voltages~\cite{Nitta1997} the presented system offers also the possibility for 
experimentally steering and switching the spin current direction.

\section{Future directions}
\label{sec:outlook}

In the present work we outlined general theoretical and
computational concepts of coherent spin-dependent transport at low temperatures
and focussed, with regard to numerical examples and possible experimental 
realizations, onto ballistic two-dimensional nanostructures based on 
non-magnetic high-mobility semiconductors. 

In order to experimentally
achieve high spin polarizations and reasonable spin currents,
if possible at room temperature, broad efforts are made to
investigate and design novel materials for spintronics.
Here, prominent and promising examples, both with respect to
fundamental physics and possible applications, are magnetic 
semiconductors such as GaMnAs \cite{Ohno1996} or semimagnetic 
materials with huge $g$-factors, for instance HgTe \cite{Koenig2007}.
Charge transport in these materials is based on holes. However,
relatively few theoretical papers deal with phase coherence effects
for hole (spin) transport, though the rich band structure and the
interplay between heavy and light hole (or electron- and hole-like)
degree of freedoms promise interesting additional features.

The theoretical methods for quantum transport, presented here 
in the context of mesoscopic systems, are also applied and extended
to treat transport in a further prospective field, namely through 
single-molecule junctions \cite{CFR06}, 
for instance (break) junctions with a molecule bridging the gap
between two leads or scanning tunneling microscope measurements
of tunnel current through molecules at surfaces. In 
{\em Molecular Spintronics}~\cite{Emberly2002,Rocha2006}
spin effects in transport through molecules are addressed. 
This subfield of spin electronics is still in its infancy. On the
computational side these systems pose considerable problems since
an adequate approach requires an appropriate description of 
the electronic and possibly vibrational properties of the molecule 
including the coupling to and effects of the leads. Whether (spin) DFT 
calculations for such an embedded molecule, combined with Landauer-type 
transport calculations, are appropriate, remains to be an issue, in
particular if charging or non-equilibrium effects are involved.

As a further future direction we expect that spin transport in graphene, 
monolayers of graphite, 
may evolve as another future research line. After its experimental discovery 
in 2004~\cite{Novoselov2004}, graphene has gained much experimental and huge theoretical 
attention owing to its many exotic properties such as the massless charge 
carriers, internal spin-like degree of freedoms and unconventional
transport characteristics. Also first experiments on graphene-based 
nanoconductors, e.g., measurements of the Aharonov-Bohm effect in graphene rings
\cite{Russo2007}, are on their way. Graphene is also viewed as a prospective 
candidate for spin-electronics, since the spin decoherence and spin relaxation times
in graphene are expected to be long~\cite{Huertas-Hernando2006,Min2006}. Recent
promising experiments already succeeded in injecting spins from ferromagnetic 
metallic contacts into graphene, although the conductance mismatch between 
graphene and the ferromagnetic leads is expected to suppress the efficiency. 
Recent theoretical proposals predict efficient spin injection into bulk graphene
from graphene ribbons employing the occurrence of current-carrying spin-polarized 
edge states in the ribbons \cite{Wimmer2007}.

\section{Acknowledgements}

Parts of the results presented are based on previous work together
with D.~Bercioux, D.~Frustaglia, and A.~Pfund.
We further thank I.~Adagideli, M.~Grifoni and D.~Weiss
for useful discussions. Two of us, KR and MW, acknowledge 
financial support by the Deutsche Forschungsgemeinschaft 
(SFB 689) and MS through the Studienstiftung des Deutschen Volkes.


\end{document}